\documentclass[aps,pra,reprint, amsmath, amssymb,superscriptaddress]{revtex4-1}
 
\usepackage{bm}
\usepackage[retainorgcmds]{IEEEtrantools}
\usepackage{graphicx}
\usepackage{mathrsfs}
\usepackage{amsmath}
\usepackage{amsfonts}
\usepackage{amssymb}
\usepackage{color}
\usepackage{times,txfonts}
\usepackage{nicefrac}
\usepackage{ragged2e}
\usepackage{tikz}

\usepackage[colorlinks=true,linkcolor=blue,urlcolor=blue,citecolor=blue,pdfusetitle]{hyperref}

\usepackage{blindtext}

\newcommand{\tr}{\mathrm{tr}}

\definecolor{cadmiumgreen}{HTML}{097969}

\begin{document}

\title{Waiting-times statistics  in boundary driven free fermion chains}
\date{\today}
\author{Gabriel T. Landi}
\email{gtlandi@gmail.com}
\affiliation{Instituto de F\'isica da Universidade de S\~ao Paulo,  05314-970 S\~ao Paulo, Brazil.}
\affiliation{School of Physics, Trinity College Dublin, College Green, Dublin 2, Ireland}
\begin{abstract}

We study the waiting-time distributions (WTDs) of quantum chains coupled to two Lindblad baths at each end. 
Our focus is on free fermion chains, where we derive closed-form expressions {\color{black}in terms of single-particle matrices}, allowing one to study arbitrarily large chain sizes. In doing so, we also derive formulas for 2-point correlation functions involving non-Hermitian propagators. 

\end{abstract}

\maketitle{}

\section{Introduction}


Transport in quantum chains constitutes a major research direction in non-equilibrium physics. 
The interplay between quantum coherent interactions and dissipative elements is known to produce a wide variety of physical phenomena. 
The basic example is the tuning of the ensuing transport regimes (e.g. ballistic, diffusive, etc.), which can be accomplished e.g. by modifying the internal system interaction~\cite{Bertini2020,Znidaric2011,Landi2015a,Gopalakrishnan2019,Bulchandani2019,Ilievski2018}. 
Further tuning the dissipation can also lead to  noise-enhanced transport~\cite{Viciani2015,Plenio2008,Biggerstaff2016,Maier2019,DeLeon-Montiel2015,Dwiputra2020}. 
These developments open the prospect for numerous potential applications, such as quantum thermoelectricity~\cite{Benenti2017a,Mahan1996,Yamamoto2015a,Dubi2011,Whitney2014} and thermal rectifiers~\cite{Li2012,Pereira2013a,Werlang2014,Avila2013,Schuab2016a,Balachandran2018,Pereira2010b,Pereira2010,Wang2007,Hu2006,Landi2014b,Silva2020,Chioquetta2021}. 

As far as transport is concerned, most studies in quantum chains focus on either one of two scenarios~\cite{Bertini2020}. 
The first is unitary time evolution, where the system is prepared in a localized  wave-packet and is then allowed to evolve unitarily. 
And the second is the steady-state that is obtained when the system is placed in contact with two baths at different temperatures and/or chemical potentials. 
This is further divided into systems described in terms of coherent transport, e.g.  the Landauer-B\"uttiker formalism~\cite{Datta1997a,Benenti2017a}, or systems described in terms of a quantum master equation, often referred to as boundary-driven systems~\cite{Landi2021}.

{\color{black}In the case of steady-states, even though the density matrix is no longer changing in time, the underlying process is still stochastic:}
At any given time, an excitation may enter from one of the baths and then travel through the system (possibly interacting with other excitations) until it eventually leaves to either bath.
The quantum nature of the system makes this description much richer, as interference effects abound. 
{\color{black}But if one only looks at the  steady-state density matrix, these effects are completely ignored.}

\begin{figure}
    \centering
    \includegraphics[width=0.5\textwidth]{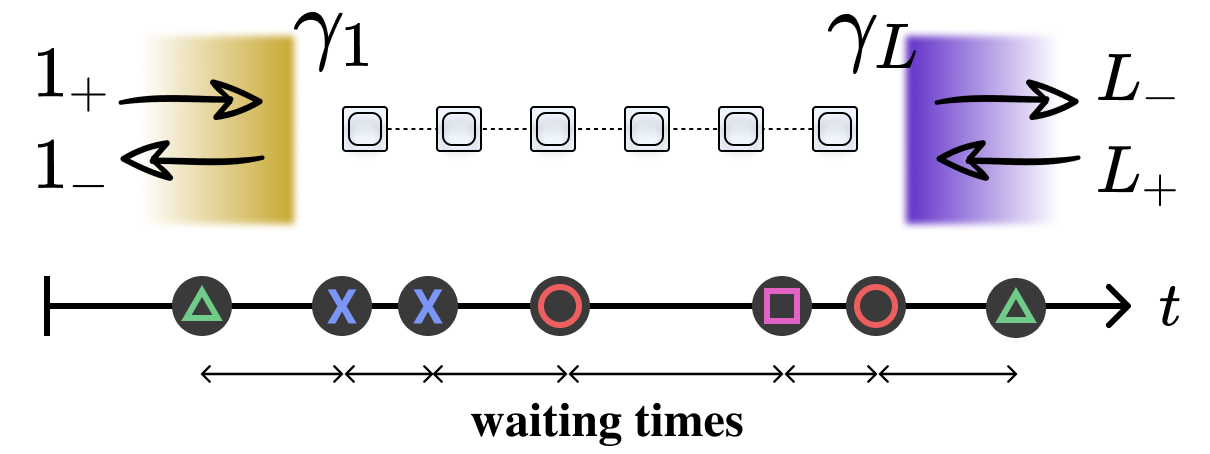}
    \caption{Top: A quantum chain of length $L$, with four possible dissipation channels, associated to the injection/extraction of an excitation at the first/last sites, with coupling strengths $\gamma_{1(L)}$. 
    Bottom: the interest in this work is on the waiting time distribution between clicks in each channel (here represented by buttons of a video game controller).}
    \label{fig:drawing}
\end{figure}

The problem can be viewed pictorially as a detector with four colors, representing an excitation entering/leaving the left/right baths (Fig.~\ref{fig:drawing}). 
Each time an event occurs, a certain color clicks. 
The complete statistics of the detection events, including the times between clicks, as well as the colors of the clicks, is captured by the theory of Full Counting Statistics (FCS)~\cite{Levitov1993,Esposito2007,Esposito2009,Brandes2008}.
The toolbox of FCS is extremely powerful, but usually difficult to apply, specially on many-body systems. 
For this reason, most practical studies on FCS have focused on the long-time statistics; i.e., on the accumulated number of clicks after a very long time, which satisfies a large-deviation principle~\cite{Touchette2009,Touchette2012}. 

A particularly interesting aspect of FCS concerns the waiting time distribution (WTD) between successive clicks~\cite{Cohen-Tannoudji1986,Plenio1998a}. 
There has been significant work on the study of WTDs in coherent conductors~\cite{Brandes2008,Schaller2009,Albert2011,Albert2012,Rajabi2013,Thomas2013,Thomas2014,Haack2014,Dasenbrook2015,Ptaszynski2017a,Ptaszynski2017,Stegmann2021,Stegmann2018}, such as double quantum dots or point contacts.
However, WTDs {\color{black}are also useful  in} many other problems, where they have not yet been thoroughly explored.
This manuscript will be concerned with boundary driven systems, comprised of a one-dimensional quantum chain coupled to two baths at each end, as described by a Lindblad master equation.
The theory of WTDs in this case was laid down in~\cite{Brandes2008}, and subsequently applied to double quantum dot systems~\cite{Schaller2009,Ptaszynski2017a}, Cooper pair splitters~\cite{Walldorf2018} and synchronized charge oscillations~\cite{Kleinherbers2021}.

{\color{black}Here} we develop formulas for the waiting-time distribution of free fermion chain. 
As with most non-interacting problems, this allows the WTD to be written in terms of matrix elements and determinants of $L\times L$ matrices (where $L$ is the number of sites in the chain), hence allowing one to study chains of arbitrary size. 
Despite being a non-interacting problem, the analysis turns out to be non-trivial since the time evolution between quantum jumps is non-Hermitian~\cite{Wiseman2009}. 
For this reason, we proceed by first casting the WTDs in terms of 2-point correlation functions involving non-Hermitian unitary evolution operators.
We then develop general formulas for such propagators, which could find use beyond the present context. 
As an application, we study a simple tight-binding chain.

\section{Formal framework}

We consider a one-dimensional fermionic chain with $L$ sites, each represented by an annihilation operator $c_i$. 
The system Hamiltonian is assumed to be quadratic, of the form
\begin{equation}\label{H}
    H = \sum\limits_{i,j} h_{ij} c_i^\dagger c_j,
\end{equation}
with a coefficient matrix $h$. 
The WTDs of free fermion chains were studied in~\cite{Thomas2014}, but only in the case of unitary dynamics.
Instead, here we assume the system evolves connected to two local baths, coupled at sites 1 and $L$, and kept at Fermi-Dirac distributions $f_1$ and $f_L$. 
The dynamics is thus assumed to be governed by the local master equation
\begin{equation}\label{M}
    \frac{d\rho}{dt} = \mathcal{L}(\rho) =  - i [H,\rho] + \sum\limits_{i = 1,L} \Big\{\gamma_i^- D[c_i] + \gamma_i^+ D[c_i^\dagger]\Big\},
\end{equation}
where $\gamma_i^- = \gamma_i (1-f_i)$ and $\gamma_i^+ = \gamma_i f_i$, with $\gamma_i$ being the coupling strengths to each bath. 
{\color{black}Here $D[A] = A \rho A^\dagger - \frac{1}{2}\{A^\dagger A, \rho\}$ is a Lindblad dissipator with arbitrary operator $A$.}

{\color{black}The WTD in this case is defined in} the context of Full Counting Statistics.
We split the Liouvillian in Eq.~\eqref{M} as
\begin{equation}
    \mathcal{L} = \mathcal{L}_0 + \sum\limits_k \mathcal{J}_k,
\end{equation}
where $\mathcal{J}_k$ represent the four possible jump channels (``four colors in the detector''), which we label as  $1_-$, $1_+$, $L_-$, $L_+$:
\begin{IEEEeqnarray}{rCCCCCL}
\mathcal{J}_{1_-}(\rho) &=& \gamma_1^- c_1 \rho c_1^\dagger
&\qquad & 
\mathcal{J}_{1_+}(\rho) &=& \gamma_1^+ c_1^\dagger \rho c_1
\nonumber \\ \label{jumps} \\\nonumber
\mathcal{J}_{L_-}(\rho) &=& \gamma_L^- c_L \rho c_L^\dagger
&\qquad & 
\mathcal{J}_{L_+}(\rho) &=& \gamma_L^+ c_L^\dagger \rho c_L
\end{IEEEeqnarray}
For instance, channel $L_-$ means an excitation was {\color{black}absorbed by} the right bath (at site $L$), and so on.

Starting from an arbitrary state $\rho$, the WTD between a jump in channel $q$ at time $0$ and a jump in channel $k$ at time $t$ is then given by~\cite{Brandes2008}:
\begin{equation}\label{WTD}
    P(t,k|q) = \frac{\tr~ \mathcal{J}_k e^{\mathcal{L}_0 t} \mathcal{J}_q (\rho)}{\tr~ \mathcal{J}_q (\rho)},
\end{equation}
which {\color{black}is} normalized as 
\begin{equation}\label{WTD_normalization}
    \sum\limits_k \int\limits_0^\infty  P(t,k|q)  dt = 1, \qquad \forall q.
\end{equation}
Eq.~\eqref{WTD} is a {\color{black}(conditional)} joint distribution representing both the time between clicks, as well as the channel of the click (note that clicks from different channels are usually statistically correlated~\cite{Dasenbrook2015}).

The {\color{black}marginal} probability that jump $q$ is followed by jump $k$, irrespective of when it occurs, is 
\begin{equation}\label{p_channel}
    p(k|q) = \int\limits_0^\infty P(t,k|q)~dt.
\end{equation}
We can also filter the WTD to consider only the statistics conditioned on the sequence of jumps being $q \to k$.
From Bayes's rule one has:
\begin{equation}\label{WTD_kq}
    P(t|k,q) = P(t,k|q)/p(k|q), 
\end{equation}
This is now a {\color{black} properly normalized WTD}, and so is more suitable for computing expectation values. 
{\color{black}We denote by $T$ the random waiting time between any two events.}
{\color{black}The average $E(T|k,q)$, conditioned on the sequence of channels} $q\to k$, is  
\begin{equation}\label{ave}
    E(T|k,q) = \int\limits_0^\infty ~t~P(t|k,q)~dt.
\end{equation}
Similarly, the variance of the waiting time reads
\begin{equation}\label{variance}
    {\rm var}(T|k,q) = E(T^2|k,q) - E(T|k,q)^2, 
\end{equation}
where $E(T^2|k,q)$ is defined similarly as $E(T|k,q)$. 

We call attention to the fact that the WTDs defined above assume that all four channels are constantly being monitored (called ``exclusive'' WTDs in~\cite{Walldorf2018}). 
One could also study a situation where only channel $k$ is being monitored (``inclusive'' WTD). 
Unfortunately, this is not  related to~\eqref{WTD} in a simple way, since the inclusive distribution must account for all possible jumps to the other channels before a click in $k$ is detected. 

The WTD~\eqref{WTD} refers to specific channels $q \to k$. 
One may also be interested in what shall be referred to as the  \emph{net activity time distribution (NATD)}, which is the WTD between any two events, irrespective of the channel. 
{\color{black}In the steady-state, it can be defined as }
\begin{equation}\label{NATD}
    P(t) = \sum\limits_{k,q} P(t,k|q) p(q), 
\end{equation}
where $p(q)$ is the relative frequency of occurrence for a jump of type $q$ (in the steady-state) and 
is given, up to a normalization, by
$p(q) = \tr \mathcal{J}_q \rho$.
{\color{black}Expectation values for NATDs may be defined similarly to e.g. Eqs.~\eqref{ave} and~\eqref{variance}, and will be denoted by $E(T)$, ${\rm var}(T)$, etc. }


Computing the waiting time distribution is generally hard, as it involves studying the evolution under the map $\mathcal{L}_0$, which is {\color{black}generally not completely positive and trace preserving}.
In fact, $\mathcal{L}_0$ can be decomposed as 
$\mathcal{L}_0 = -i (H_e \rho - \rho H_e^\dagger)$, where
\begin{IEEEeqnarray}{rCL}
\label{He}
    H_e &=& H - \frac{i}{2} \Bigg[ \gamma_1 (1-f_1) c_1^\dagger c_1  + \gamma_1 f_1 c_1 c_1^\dagger,
    \\[0.2cm]
    &&\qquad + \gamma_L (1-f_L) c_L^\dagger c_L + \gamma_L f_L c_L c_L^\dagger 
    \Bigg].\nonumber
\end{IEEEeqnarray}
{\color{black}Hence, the action of $\mathcal{L}_0$ is tantamount to a non-Hermitian Hamiltonian evolution.}
Given the four possible channels in Eq.~\eqref{jumps}, there can be in total 16 WTDs~\eqref{WTD}. 
They can be more compactly written as 
\begin{IEEEeqnarray}{rCl}
\label{mp}
    P(t, i_-| j_+) &=& \frac{\gamma_i^-}{\langle c_j c_j^\dagger \rangle} \tr\Big\{ c_i^\dagger c_i e^{-i H_e t} c_j^\dagger \rho c_j e^{i H_e^\dagger t}\Big\},
    \\[0.2cm]
\label{pp}    
    P(t, i_+| j_+) &=& \frac{\gamma_i^+}{\langle c_j c_j^\dagger \rangle} \tr\Big\{ c_i c_i^\dagger e^{-i H_e t} c_j^\dagger \rho c_j e^{i H_e^\dagger t}\Big\},
    \\[0.2cm]
\label{mm}    
    P(t, i_-| j_-) &=& \frac{\gamma_i^-}{\langle c_j^\dagger c_j \rangle} \tr\Big\{ c_i^\dagger c_i e^{-i H_e t} c_j \rho c_j^\dagger e^{i H_e^\dagger t}\Big\},
    \\[0.2cm]
    P(t, i_+| j_-) &=& \frac{\gamma_i^+}{\langle c_j^\dagger c_j \rangle} \tr\Big\{ c_i c_i^\dagger e^{-i H_e t} c_j \rho c_j^\dagger e^{i H_e^\dagger t}\Big\},
\label{pm}    
\end{IEEEeqnarray}
with $i,j = 1,L$.

\section{Trace-det formulas for non-Hermitian fermionic forms}

The traces in the WTDs~\eqref{mp}-\eqref{pm} resemble 2-time correlation functions.
However, the biggest difference is that the time propagator is $H_e$, which is non-Hermitian. 
This makes the direct computation of the WTDs  more difficult than they may seem at first. 
For instance, one cannot use the usual Baker–Campbell–Hausdorff formulas~\cite{Logan2005}, since the quantities in question here are of the form 
$e^{-i H_e t} \mathcal{O} e^{i H_e^\dagger t}$, instead of 
$e^{-i H_e t} \mathcal{O} e^{i H_e t}$. 
Instead, to compute these traces, we first develop a series of formulas which hold even for non-Hermitian operators. 
They are all based on variations of the so-called 
Blankenbecler-Scalapino-Sugar (BSS)
``trace-det'' relations~\cite{Blankenbecler1981,Klich2014}, which are widely used in quantum Monte Carlo.  
{\color{black}Below, we only provide an overview of the main results. The actual derivations are given in  Appendix~\ref{app:formula}. }

Let 
$\mathcal{X} = \sum_{ij} X_{ij} c_i^\dagger c_j$, 
$\mathcal{Y} = \sum_{ij} Y_{ij} c_i^\dagger c_j$,
$\mathcal{Z} = \sum_{ij} Z_{ij} c_i^\dagger c_j$
be quadratic forms in fermionic operators, with  arbitrary coefficient matrices $X$, $Y$ and $Z$. 
The BSS trace-det formula states that~\cite{Blankenbecler1981,Klich2014}
\begin{equation}\label{tr_det}
    \tr \big\{e^{\mathcal{X}} e^{\mathcal{Y}}e^{\mathcal{Z}} \big\} = \det (1 + e^X e^Y e^Z). 
\end{equation}
Here and henceforth we will not distinguish between the number 1 and the identity matrix 1.
Eq.~\eqref{tr_det} extends identically to more than three operators; but for our purposes 3 will suffice.
This formula provides a huge simplification since the rhs is a determinant on the space of $L\times L$ matrices, $X$, $Y$, $Z$. 
This is to be contrasted with the lhs, which is a trace of a $2^L \times 2^L$ dimensional operator. 

Using Eq.~\eqref{tr_det}, we show in Appendix~\ref{app:formula} that 
\begin{equation}\label{tr_det_cdc}
    \tr \big\{ c_i^\dagger c_{i'} ~e^{\mathcal{X}}e^{\mathcal{Y}}e^{\mathcal{Z}} \big\} 
    = \mathbb{D} \mathcal{T}_{i'i},
\end{equation}
where 
\begin{equation}\label{Dmat}
    \mathbb{D} = \det (1 + e^X e^Y e^Z),
\end{equation}
and 
\begin{equation}\label{Tmat}
    \mathcal{T} = ( e^{-Z} e^{-Y} e^{-X} + 1)^{-1} = e^X e^Y e^Z (1 + e^X e^Y e^Z)^{-1}.
\end{equation}
Eq.~\eqref{tr_det_cdc} again holds for more than 3 exponentials, provided the order of the exponentials  are preserved. 

\begin{widetext}
Similarly, using both~\eqref{tr_det} and~\eqref{tr_det_cdc}, we show in Appendix~\ref{app:formula} that 
\begin{IEEEeqnarray}{rCl}\label{tr_det_2_cs}
\tr \Big\{ c_i^\dagger c_{i'} e^{\mathcal{X}} c_j^\dagger c_{j'} e^{\mathcal{Y}}e^{\mathcal{Z}} \Big\} &=&\mathbb{D}\Bigg[
    (e^{-X} \mathcal{T} e^X)_{j'j} \mathcal{T}_{i'i}
    +(\mathcal{T} e^{-Z} e^{-Y})_{i'j} (e^{-X} \mathcal{T})_{j'i} \Bigg].
\end{IEEEeqnarray}
Compared with, e.g., Eq.~\eqref{mp}, the main difference is that here there is a term $c_j^\dagger c_{j'} e^{\mathcal{Y}}$ while in~\eqref{mp} it reads 
$c_j^\dagger e^{\mathcal{Y}} c_{j'}$.
Using the fact that $e^{\mathcal{Y}} c_i e^{-\mathcal{Y}} = \sum_j (e^{Y})_{ij} c_j$, together with the fermionic algebra, one finds that 
\begin{IEEEeqnarray}{rCl}\label{2cs_mp}
\tr \Big\{ c_i^\dagger c_{i'} e^{\mathcal{X}} c_j^\dagger  e^{\mathcal{Y}} c_{j'} e^{\mathcal{Z}} \Big\} &=&\mathbb{D}\Bigg[
    (e^{-Y} e^{-X} \mathcal{T} e^X)_{j'j} \mathcal{T}_{i'i}
    + (\mathcal{T} e^{-Z} e^{-Y})_{i'j} (e^{-Y} e^{-X} \mathcal{T})_{j'i} \Bigg].
\end{IEEEeqnarray}
This is of the same form as the trace appearing in Eq.~\eqref{mp}, provided we take $i'=i$ and $j' = j$. 
Proceeding similarly, we can also compute expressions for the other 3 traces in Eqs.~\eqref{pp}-\eqref{pm}:
\begin{IEEEeqnarray}{rCl}
\label{2cs_pp}
\tr \Big\{ c_i c_{i'}^\dagger e^{\mathcal{X}} c_j^\dagger  e^{\mathcal{Y}} c_{j'} e^{\mathcal{Z}} \Big\} &=&\mathbb{D}\Bigg[
    (e^{-Y} e^{-X} \mathcal{T} e^X)_{j'j} \big(\delta_{ii'} - \mathcal{T}_{ii'}\big)
    - (\mathcal{T} e^{-Z} e^{-Y})_{ij} (e^{-Y} e^{-X} \mathcal{T})_{j'i'} \Bigg],
    \\[0.2cm]
\label{2cs_mm}
\tr \Big\{ c_i^\dagger c_{i'} e^{\mathcal{X}} c_j^\dagger  e^{\mathcal{Y}} c_{j'} e^{\mathcal{Z}} \Big\} &=&\mathbb{D}\Bigg[
    \Big[ (e^Y)_{jj'} - (e^{-X} \mathcal{T} e^X e^Y)_{j'j} \Big] \mathcal{T}_{i'i}
    - (\mathcal{T} e^{-Z})_{i'j'} (e^{-X} \mathcal{T})_{ji} \Bigg],
    \\[0.2cm]
\label{2cs_pm}
\tr \Big\{ c_i^\dagger c_{i'} e^{\mathcal{X}} c_j e^{\mathcal{Y}} c_{j'}^\dagger  e^{\mathcal{Z}} \Big\} &=&\mathbb{D}\Bigg[
    \Big[ (e^Y)_{jj'} - (e^{-X} \mathcal{T} e^X e^Y)_{jj'} \Big]\big(\delta_{ii'}- \mathcal{T}_{ii'}\big)
    + (\mathcal{T} e^{-Z})_{ij'} (e^{-X} \mathcal{T})_{ji'} \Bigg],
\end{IEEEeqnarray}
All formulas hold for \emph{arbitrary} matrices $X,Y,Z$.
But before we can apply them to the WTDs, some adaptations are still required. 
\end{widetext}


\section{Computation of the WTD{\scriptsize{s}}}

Since $H_e$ in Eq.~\eqref{He} is a quadratic form, we can use 
Eqs.~\eqref{2cs_mp}-\eqref{2cs_pm} to compute the WTDs~\eqref{mp}-\eqref{pm} provided the initial state $\rho$ is Gaussian. We will focus on two main choices of initial states:  the steady-state $\rho_{\rm ss}$ of the master equation~\eqref{M} and the vacuum state $\rho_{\rm vac} = |0\rangle\langle 0 |$. 
We can consider both together, by taking a generic Gaussian initial state of the form 
\begin{equation}\label{rho_gaussian}
    \rho = \frac{1}{\mathbb{Z}}e^{- \sum\limits_{ij} M_{i,j} c_i^\dagger c_j},
\end{equation}
with some $L\times L$ matrix $M$. The partition function $\mathbb{Z}$ is, in light of Eq.~\eqref{tr_det},
\begin{equation}\label{Z}
    \mathbb{Z} = \det (1 + e^{-M}).
\end{equation}
Alternatively, one can also characterize the Gaussian state by the covariance matrix $C_{ij} = \langle c_j^\dagger c_i \rangle$. 
The relation between $C$ and $M$ reads
\begin{equation}\label{CM}
    e^M = \frac{1-C}{C},
    \qquad 
    C = \frac{1}{e^M+1}.
\end{equation}

The quadratic nature of the master equation~\eqref{M} implies that the standard time evolution of $C$ will be given by a Lyapunov equation 
\begin{equation}\label{lyapunov}
    \frac{dC}{dt} = - (W C + C W^\dagger) + F. 
\end{equation}
where
\begin{IEEEeqnarray}{rCl}
\label{W}
W &=& i h + \frac{1}{2} {\rm diag}\big(\gamma_1, 0,\ldots,0,\gamma_L\big), 
\\[0.2cm]
F &=& {\rm diag} \big(\gamma_1 f_1, 0,\ldots,0,\gamma_L f_L\big).
\label{F}
\end{IEEEeqnarray}
The steady-state is thus the long-time solution of Eq.~\eqref{lyapunov}; viz., 
\begin{equation}\label{lyapunov_ss}
    W C_{\rm ss} + C_{\rm ss} W^\dagger = F. 
\end{equation}
Similarly, the vacuum state is simply $C_{\rm vac} = 0$. {\color{black}
In practice, it may be more convenient to set $C_{\rm vac}$ to be proportional to the identity, with some small constant that is ultimately taken to zero. This approach will actually be used below, around Eq.~\eqref{DTmat_WMW}}.
One should also bear in mind that the conditional evolution which appears in the WTDs, is \emph{not} Gaussian because e.g. $c_q \rho c_q^\dagger$ is not a Gaussian state. 
Notwithstanding, as we will show, it is still possible (and convenient) to express most results in terms of the matrices $C$ (or $M$), $W,F$.


The operator $H_e$ in Eq.~\eqref{He} is not yet in a canonical quadratic form due to the terms $c_1 c_1^\dagger$ and $c_L c_L^\dagger$. 
{\color{black}In fact, writing} $c_k c_k^\dagger = 1 - c_k^\dagger c_k$ turns out to yield a non-trivial constant. 
The resulting Hamiltonian can be conveniently written as
\begin{equation}\label{He_quad}
    H_e = -i \sum\limits_{i,j} Q_{ij} c_i^\dagger c_j - \frac{i}{2} \Gamma \equiv \tilde{H}_e - \frac{i}{2} \Gamma, 
\end{equation}
where $\Gamma = \gamma_1 f_1 + \gamma_L f_L$ is a constant and
\begin{equation}\label{quelle}
Q = W - F,
\end{equation}
A trace such as that in Eq.~\eqref{mp} can thus {\color{black}finally} be written as 
\begin{equation}
    \tr\Big\{ c_i^\dagger c_i e^{-i H_e t} c_j^\dagger \rho c_j e^{i H_e^\dagger t}\Big\} = \frac{e^{-\Gamma t} }{\mathbb{Z}} 
    \tr\Big\{ c_i^\dagger c_i e^{-i \tilde{H}_e t} c_j^\dagger e^{-\sum_{k\ell} M_{k\ell} c_k^\dagger c_\ell} c_j e^{i \tilde{H}_e^\dagger t}\Big\},
\end{equation}
which is now in the form~\eqref{2cs_mp}, provided we identify 
\begin{equation}\label{XYZ_WMW}
    X = -Q t, \qquad Y = - M, \qquad Z = -Q^\dagger t.
\end{equation}
The final expression for all WTDs therefore reads

\begin{widetext}
\begin{IEEEeqnarray}{rCl}
\label{mp_final}
    P(t,i_-|j_+) &=& \frac{\gamma_i^-}{1-C_{jj}}\frac{e^{-\Gamma t}}{\mathbb{Z}}\mathbb{D} 
    \Bigg\{
    (e^{M} e^{Qt} \mathcal{T} e^{-Q t})_{jj} \mathcal{T}_{ii}
    +
     (\mathcal{T} e^{Q^\dagger t} e^{M})_{ij} (e^{M} e^{Q t} \mathcal{T})_{ji} \Bigg\},
\\[0.2cm]
\label{pp_final}
    P(t,i_+|j_+) &=& \frac{\gamma_i^+}{1-C_{jj}}\frac{e^{-\Gamma t}}{\mathbb{Z}}\mathbb{D} 
    \Bigg\{
    (e^{M} e^{Qt} \mathcal{T} e^{-Q t})_{jj} \big(1-\mathcal{T}_{ii}\big)
    -
     (\mathcal{T} e^{Q^\dagger t} e^{M})_{ij} (e^{M} e^{Q t} \mathcal{T})_{ji} \Bigg\},
\\[0.2cm]     
\label{mm_final}
    P(t,i_-|j_-) &=& \frac{\gamma_i^-}{C_{jj}}\frac{e^{-\Gamma t}}{\mathbb{Z}}\mathbb{D} 
    \Bigg\{
    \Big[(e^{-M})_{jj} - 
    (e^{Qt} \mathcal{T} e^{-Q t} e^{-M})_{jj}\Big] \mathcal{T}_{ii}
    -
     (\mathcal{T} e^{Q^\dagger t} )_{ij} (e^{Q t} \mathcal{T})_{ji} \Bigg\},
\\[0.2cm]     
\label{pm_final}
    P(t,i_+|j_-) &=& \frac{\gamma_i^+}{C_{jj}}\frac{e^{-\Gamma t}}{\mathbb{Z}}\mathbb{D} 
    \Bigg\{
    \Big[(e^{-M})_{jj} - 
    (e^{Qt} \mathcal{T} e^{-Q t} e^{-M})_{jj}\Big] \big(1-\mathcal{T}_{ii}\big)
    +
     (\mathcal{T} e^{Q^\dagger t} )_{ij} (e^{Q t} \mathcal{T})_{ji} \Bigg\},     
\end{IEEEeqnarray}
where
\begin{equation}\label{DTmat_WMW}
    \mathbb{D} = \det \big(1 + e^{-Q t} e^{-M} e^{-Q^\dagger t}\big), \qquad 
    \mathcal{T} = \big( e^{Q^\dagger t} e^{M} e^{Q t} + 1\big)^{-1}.
\end{equation}
In view of the fact that $e^M = (1-C)/C$, we therefore see that everything is expressed in terms of {\color{black}the} quantities $C,W,F$ associated to the Lyapunov equation~\eqref{lyapunov}, which is nice.
\end{widetext}

Next we specialize these formulas to the case where the initial state is the vacuum, $C_{\rm vac} = 0$. 
It is prudent to first assume $C$ is proportional to the identity, $C = \lambda$, and then take $\lambda \to 0$. 
In light of Eq.~\eqref{CM}, we have that
$e^M = (1-\lambda)/\lambda$, so that in the limit $\lambda \to 0$ we get $Z = \mathbb{D} = 1$, and $C_{jj} = 0$.
Moreover, 
$\mathcal{T} = \frac{\lambda}{1-\lambda} e^{-Q t} e^{-Q^\dagger t}$.
Terms containing products of $e^M$ and $\mathcal{T}$ will thus be of order $1$, while terms containing only $\mathcal{T}$ will vanish. 
Eqs.~\eqref{mp_final}-\eqref{pp_final} thus reduce to 
\begin{IEEEeqnarray}{rCl}
\label{mp_vacuum}
    P(t,i_-|j_+) &=& \gamma_i^- e^{-\Gamma t} \Big(e^{-Q t}\Big)_{ij}
    \Big(e^{-Q^\dagger t}\Big)_{ji}.
    \\[0.2cm]
\label{pp_vacuum}    
    P(t,i_+|j_+) &=& \gamma_i^+ e^{-\Gamma t}\Big[ \big(e^{-Q^\dagger t} e^{-Qt} \big)_{jj}-\Big(e^{-Q t}\Big)_{ij}
    \Big(e^{-Q^\dagger t}\Big)_{ji}\Big].
\end{IEEEeqnarray}
The other two WTDs, Eq.~\eqref{mm_final} and~\eqref{pm_final}, vanish in this case because $c_j |0\rangle \langle 0| c_j^\dagger \equiv 0$. 


\section{Example: tight-binding model}

As an application, we consider a tight-binding model with Hamiltonian 
\begin{equation}\label{H_TB}
    H = -\sum\limits_{i=1}^L V c_i^\dagger c_i - J \sum\limits_{i=1}^{L-1} \big(c_i^\dagger c_{i+1} + c_{i+1}^\dagger c_i\big).
\end{equation}
This is a prototypical example of ballistic transport~\cite{Karevski2009,Znidaric2010a,Asadian2013}. 
We henceforth fix $V = J = 1$, $\gamma_1 = \gamma_L$, $f_1 = 1$ and $f_L = 0$. 
This means that excitations can only be \emph{injected} in site 1 or \emph{collected} on site $L$.
This reduces the problem to four WTDs, 
$P(t,L_-|1_+), P(t,1_+|1_+), P(t,1_+|L_-), P(t,L_-|L_-)$. 
Due to the symmetry $\gamma_1 = \gamma_L$, 
the first two equal the last two.
Hence, we have to focus only on 
$P(t,L_-|1_+)$ and $P(t,1_+|1_+)$. 
{\color{black}It is also important to distinguish the fundamental physical difference between these two distributions. 
Namely, $P(t,1_+|1_+)$ is a local quantity, associated to clicks on the same site, while $P(t,L_-|1_+)$ is non-local, describing  events at spatially distant points.
}

\begin{figure}
    \centering
    \includegraphics[width=0.45\textwidth]{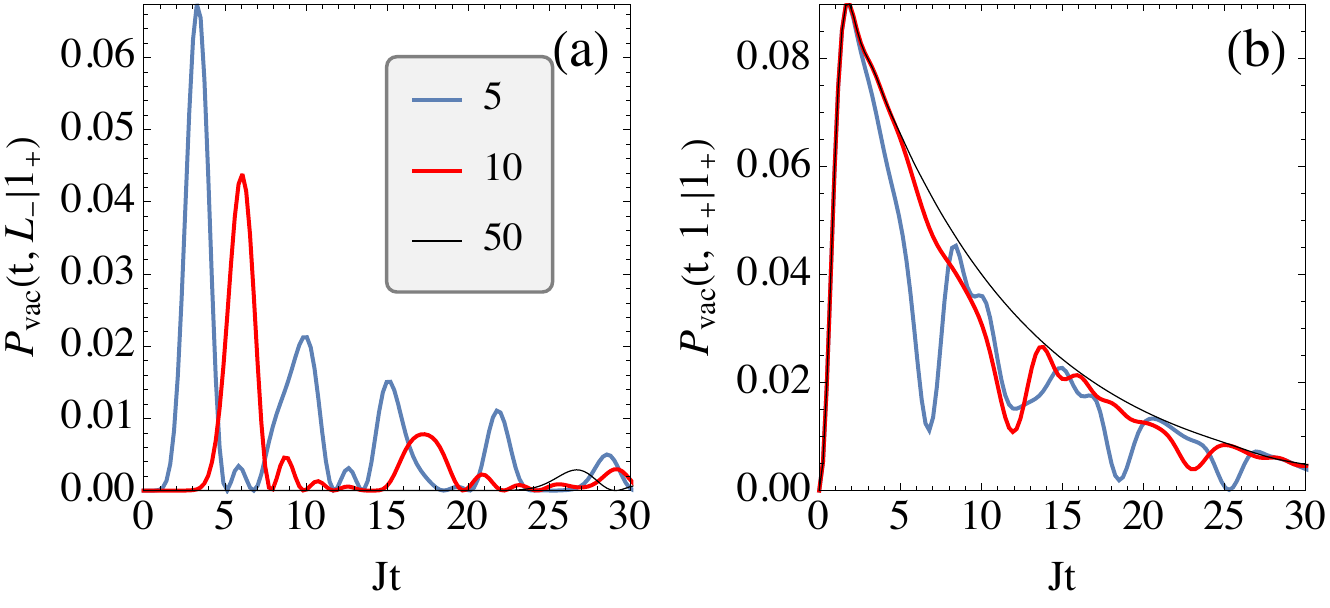}
    \caption{Waiting time distributions (a) $P_{\rm vac}(t,L_-|1_+)$ and (b) $P_{\rm vac}(t,1_+|1_+)$, starting in the vacuum. Each curve is for a different system size $L = 5,10,50$. 
    Parameters: $V = J = 1$, $\gamma_1 = \gamma_L = 0.1 J$, $f_1 = 1$, $f_L = 0$.}
    \label{fig:example_vac_wtds}
\end{figure}

\begin{figure}
    \centering
    \includegraphics[width=0.45\textwidth]{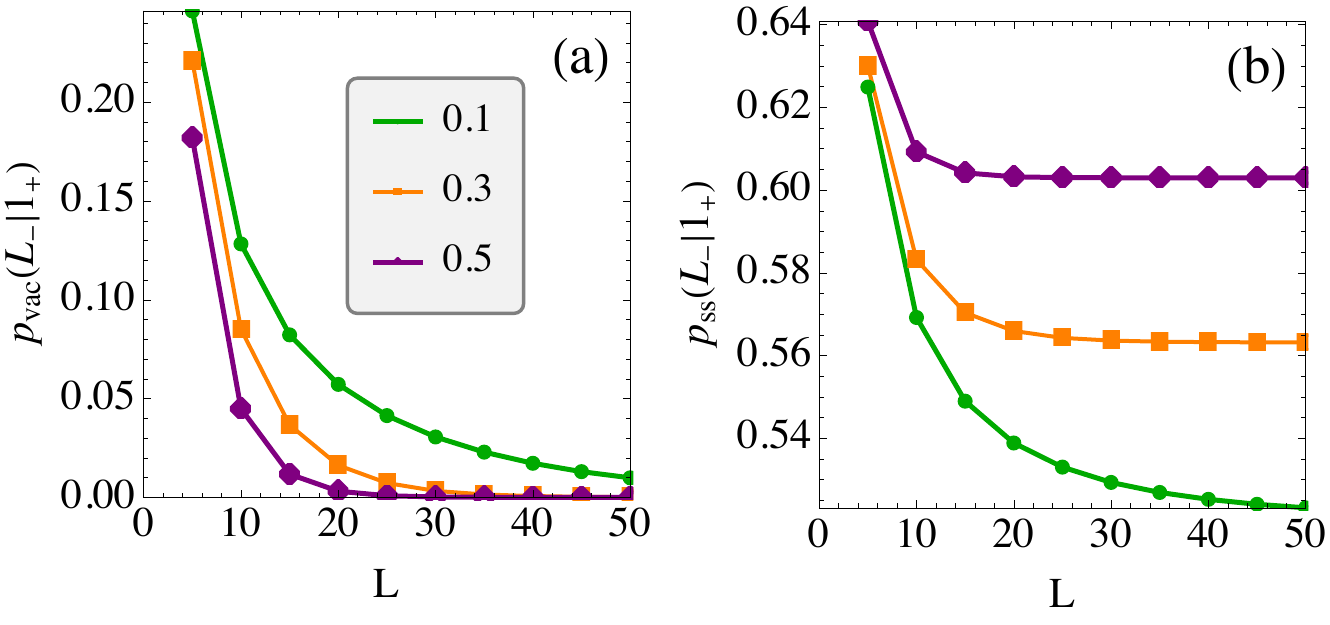}
    \caption{Detection probabilities for the right bath $p(L_-|1_+)$, as a function of the system size $L$. (a) vacuum; (b) steady-state. Each curve is for a different value of $\gamma$. Other parameters are the same as in Fig.~\ref{fig:example_vac_wtds}.}
    \label{fig:channel _probs}
\end{figure}

In analyzing these WTDs, we start by considering the case where the system is initially in the vacuum. 
The reason is that this more closely resembles standard unitary transport protocols, where a wavepacket is inserted in an empty chain, and one watches how it propagates (c.f.~Ref.~\cite{Thomas2014}). 
The corresponding WTDs are shown in Fig.~\ref{fig:example_vac_wtds}. 
The most familiar scenario is that of Fig.~\ref{fig:example_vac_wtds}(a): an excitation is created on the left and then propagates with time. 
As can be seen, the resulting WTD is initially zero {\color{black}since it takes a finite amount of time for the excitation to travel from one site to the other.}
It then presents a series of peaks, characteristic of quantum coherent processes. 
The first peak is the primary absorption, where the excitation leaves the chain (and hence a click is detected). 
The other peaks are secondary processes,
{\color{black} related to the wave-like nature of the particle's propagation in the chain, and the fact that, for finite sizes, the wavepackets may move back and forth multiple times within the chain, until they are eventually removed.}
As $L$ increases the position of the peaks tend to {\color{black}be} pushed to longer times, {\color{black} which was found from numerics to scale as $t_{\rm peak} \propto L$, exactly as expected for ballistic transport}. 
{\color{black}Moreover,} the relative magnitudes {\color{black}of the peaks  also} diminish (the curve for $L=50$ in Fig.~\ref{fig:example_vac_wtds}(a) is {\color{black} only barely visible, around $Jt \sim 27$}).
{\color{black}The reason why this happens is simply due to the way WTDs are normalized, as will be discussed further in Fig.~\ref{fig:channel _probs}.
}

\begin{figure}
    \centering
    \includegraphics[width=0.45\textwidth]{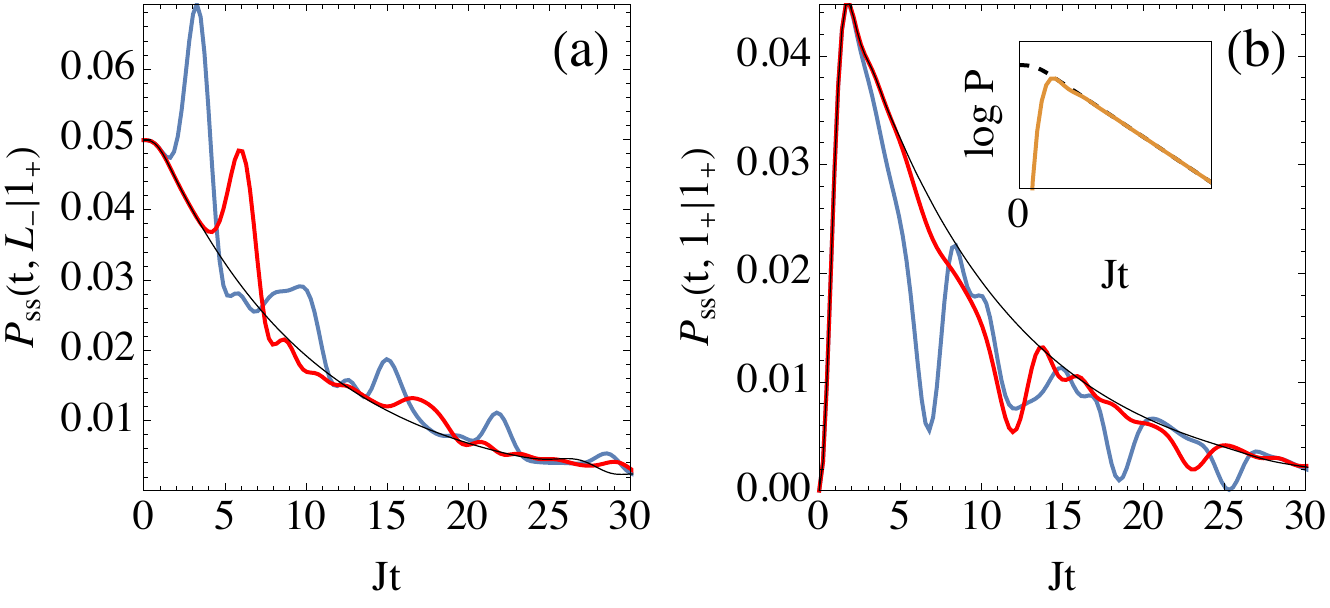}
    \caption{Similar to Fig.~\ref{fig:example_vac_wtds}, but for the system starting in the steady-state. The inset is {\color{black} a log-scale plot of $P_{\rm ss}(t,L_-|1_+)$ (black-dashed) and $P_{\rm ss}(t,1_+|1_+)$ (orange) for $L=50$. 
    }
    }
    \label{fig:example_ss_wtds}
\end{figure}

Conversely, $P(t,1_+|1_+)$, shown in Fig.~\ref{fig:example_vac_wtds}(b), is not associated to transport. 
Instead, it describes the waiting times between consecutive firings of the same channel. 
It is thus zero when $t=0$, but then rapidly {\color{black}increases}. 
The peak, which occurs at $t\sim 2$, represents the most likely waiting time between two consecutive jumps. 
When $L$ is small, the distribution presents a series of oscillations, associated to the confinement of the ejected excitation in a finite-size chain. 
But as $L$ gets large, the distribution -- and hence the spacing between firing times -- quickly becomes independent of $L$.
{\color{black}In fact, for large sizes $P(t,1_+|1_+)$ essentially follows an exponential  distribution, with a characteristic time $1/\gamma$, dictated precisely by the Lindblad coupling strength.}

Still concerning $P(t,1_+|1_+)$, Fig.~\ref{fig:example_vac_wtds}(b),
it is possible to draw an analogy with queuing theory -- i.e., the description of customers arriving in a queue. 
At any given time, the environment is sending multiple excitations to the system. 
Precisely how it does that is not an information  that is present in the master equation, only in the microscopic model of the system-environment interactions. 
As far as the master equation is concerned, however, all that matters is how many of those excitations actually enter the system. 
In queuing theory, this would be associated to the phenomenon of \emph{balking}, which is when a customer arrives at the line, but decides not to enter it \footnote{{\color{black}This analogy is limited by the fact that $P(t,1_+|1_+) = 0$ for $t = 0$.}}. 
The excitations that enter the system, are those that did not balk.
Except for finite size effects, one expects that balking should be associated mostly with the environment, as well as the system-environment \emph{boundary} (i.e., site 1). 
A related, but different concept, is \emph{reneging}, which is when a customer enters a line but decides to leave after some time. 
This would be associated with the WTD $P(t,1_-|1_+)$, which will in general depend on the whole chain. 
In this example, however, this effect is zero since we set $f_1 = 1$. 

In Fig.~\ref{fig:example_ss_wtds} we show similar results, but now for the system starting in the steady-state. Interestingly, in this case $P(t,1_+|1_+)$ is practically unaltered. 
This again corroborates {\color{black}the} idea that $P(t,1_+|1_+)$ is ultimately a property of site 1 and the environment. 
Conversely, the behavior of $P(t,1_+|1_+)$ in Fig.~\ref{fig:example_ss_wtds}(a) is  entirely different. 
First, it is maximal at $t=0$.
This occurs because, unlike the vacuum case of Fig.~\ref{fig:example_vac_wtds}, the system now already has plenty of other excitations, so that a click on the left bath is not a requirement for observing a click on the right one. 
In fact, one can see {\color{black} in Fig.~\ref{fig:example_ss_wtds}(a)} the same peaks of Fig.~\ref{fig:example_vac_wtds}(a), except that they are enveloped by a monotonically decaying distribution. 
When the size of the chain increases, the latter are rapidly suppressed, and $P(t,L_-|1_+)$ tends to a simple exponential decay (using larger values of $\gamma$ also have the tendency to suppress the oscillations).
In fact, the inset in Fig.~\ref{fig:example_ss_wtds}(b) shows a log-scale plot {\color{black} of both distributions} for $L = 50$. This makes it evident that, except for {\color{black} small deviations at early times}, the distributions are essentially given by a single exponential $P \sim e^{-t/\tau}$, with {\color{black}$\tau = 1/\gamma$. }

The relative frequency with which the jump $1_+ \to L_-$ occurs is given by $p(L_-|1_+)$, Eq.~\eqref{p_channel}. 
This is presented in Fig.~\ref{fig:channel _probs}, as a function of $L$, for both steady-state and vacuum.
When the system starts in the vacuum (Fig.~\ref{fig:channel _probs}(a)) $p(L_-|1_+)$ is exponentially suppressed with increasing $L$, for all values of $\gamma$. 
This happens because, when the chain is large, it takes a long time for an excitation to be transported to the other side.
In contrast, $1_+ \to 1_+$ {\color{black}refers to two events}  at the same site, and is thus independent of $L$. 
{\color{black}This explains why the curves in Fig.~\ref{fig:example_vac_wtds}(a) are suppressed with increasing $L$.}
Conversely, if the system starts in the steady-state (Fig.~\ref{fig:channel _probs}(b)), the probabilities tend to a finite value when $L \to \infty$. 
This means that the frequencies with which $L_-|1_+$ and $1_+|1_+$ occur remain comparable in magnitude, even in the thermodynamic limit. 

\begin{figure}
    \centering
    \includegraphics[width=0.45\textwidth]{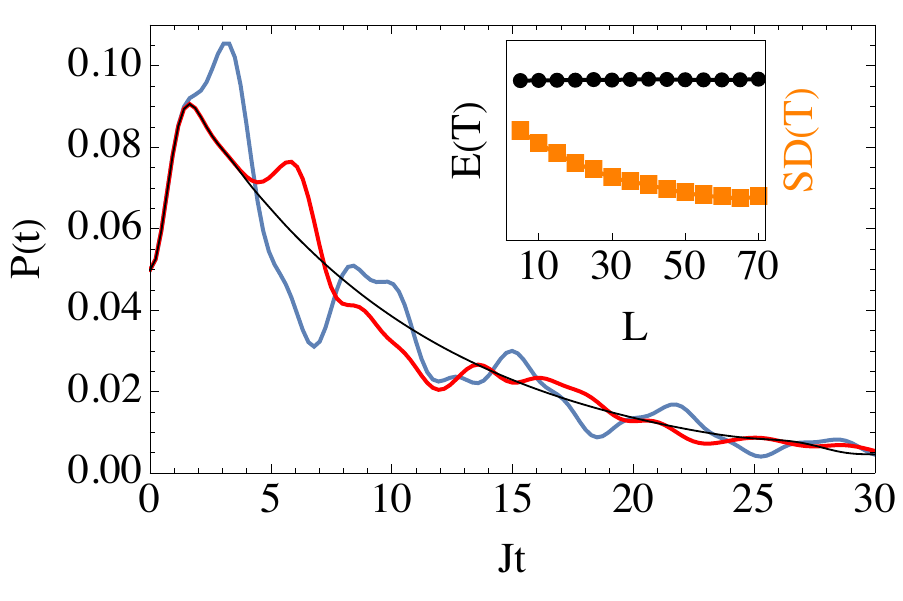}
    \caption{Activity time distribution, Eq.~\eqref{NATD}, as  a function of time for the same values of $L$ (and other parameters) as in Fig.~\ref{fig:example_vac_wtds}. 
    The inset shows the mean $E(T)$ and standard deviation ${\rm SD}(T) = \sqrt{\rm var}(T)$ as a function of the system size $L$. 
    }
    \label{fig:atd}
\end{figure}

Finally, we turn to the net activity time distribution (NATD) $P(t)$ in Eq.~\eqref{NATD}, which describes the waiting time between any two events.
The results are shown in Fig.~\ref{fig:atd}.
Due to the symmetry $P(L_-|L_-) = P(1_+|1_+)$ and $P(1_+|L_-) = P(L_-|1_+)$, of the present choice of parameters, it reduces in this case to 
$P(t) = P(t,L_-|1_+) + P(t,1_+|L_-)$.
Thus, $P(t)$ behaves as a mixture of the two distributions in Fig.~\ref{fig:example_ss_wtds}, serving as a good summary of the typical activities happening in the system. 
In the inset we show the mean and standard deviation as a function of $L$. Quite remarkably, even though the distributions themselves depend sensibly on $L$ (main plot), the mean $E(T)$ is absolutely flat. 
The standard deviation, on other hand, depends weakly on $L$ and is also very close to the mean. 

{\color{black}
To shed further light on the NATD, we look at the case $L = 2$, where it can actually be computed analytically. 
The result is 
\begin{equation}
    P(t) = \frac{\gamma}{2(\gamma^2 - 4 J^2)}e^{-\gamma t} \Bigg[\gamma^2 - 8 J^2 + 4 J^2 \cos\big(t \sqrt{4J^2 - \gamma^2}\big)\Bigg].
\end{equation}
The average time between clicks is thus 
\begin{equation}
    E(T) = \frac{1}{\gamma} + \frac{\gamma}{4J^2}.
\end{equation}
The first contribution is associated solely with the stochastic nature of the baths, which generates a typical exponential distribution with rate $\gamma$; 
it is therefore consistent with the approximate exponential behavior shown in the inset of Fig.~\ref{fig:example_ss_wtds}.
The second term, on the other hand, is associated to the coherent hoping  $J$. Hence, it yields a correction to the average waiting time due to the presence of the system.
Since $E(T)$ is found to be independent of $L$, we therefore conclude that the contributions from the hoping persist even in the thermodynamic limit.
}

\section{Significance and  applications}

The goal of this paper was to provide {\color{black}closed expressions} for the waiting time distribution in free fermion chains, {\color{black} written solely in terms of the $L\times L$ matrices characterizing the problem}. 
We believe this is of value for three reasons. 
First, WTDs represent a somewhat unexplored aspect of full counting statistics, with rich physics. 
{\color{black}For example,  in the simple tight-binding model studied in this paper we have shown how WTDs can clearly capture  dynamical aspects of transport through boundary driven chains. This includes, in particular, insights on how the chain size $L$ influences the time between absorption/emission events, the relative probabilities and the overall dynamical activities within the chain.
}
Second, WTDs are usually difficult to compute, specially for many-body systems. Being able to study them for arbitrary chain sizes is thus extremely valuable. 
For instance, they can be used to benchmark  simulations for interacting systems using, e.g., tensor networks~\cite{Mascarenhas2015,Cui2015,Werner2014,Jaschke2018,Brenes2020,Casagrande2020}. 
The third reason why these results should be of value is that, although free fermions are sometimes regarded as not so interesting (e.g., when compared to interacting models), there has recently been a surge of interest in exotic fermionic chains, such as those exhibiting quasi-periodic behavior. 
In fact, as illustrated in Refs.~\cite{Varma2017,Znidaric2017,Ganeshan2015,Znidaric2013a,Hiramoto1992,Purkayastha2017a,Chiaracane2019}, quasi-periodic non-interacting chains can exhibit any kind of transport, not only ballistic~\cite{Znidaric2013a,Varma2019,Znidaric2017,Lacerda2021}. A study of  WTDs for these models will be the subject of future work.

As for other future extensions, it would be interesting to extend this to Gaussian bosonic states, as they may have  applications in e.g., optomechanical systems~\cite{Aspelmeyer2014,Bowen2016}. 
Another extension is to include WTDs in which not all channels are monitored. 
In this case it is not possible to decompose the free evolution as $\mathcal{L}_0= -i (H_e \rho - \rho H_e^\dagger)$. 
Notwithstanding, the resulting Liouvillian is still quadratic, so it should be possible to derive the WTD  using e.g. third quantization~\cite{Prosen2008}, or a similar vectorization method~\cite{Landi2021}.

\section*{Acknowledgments}
This work was developed during quarantine at the Red Cow Moran Hotel, whose hospitality and caring from the staff was sincerely appreciated. 
The author would like to thank Gernot Schaller, for the insightful discussions on waiting times.
The author acknowledges the financial support of the S\~ao Paulo Funding Agency FAPESP (Grant No.~2019/14072-0.). 

\begin{widetext}
\appendix
\section{\label{app:formula}Proof of Eqs.~\eqref{tr_det_cdc} and~\eqref{tr_det_2_cs}}

\subsection{Proof of Eq.~\eqref{tr_det_cdc}}

We will prove Eqs.~\eqref{tr_det_cdc} and~\eqref{tr_det_2_cs} using Eq.~\eqref{tr_det}. 
For simplicity, it will be assumed that $i' = i$ and $j' = j$, but the proof when they are different is quite similar. 
Due to the fermionic algebra, it holds that for any constant $\alpha$ \footnote{If one is interested in $i'\neq i$, then the decomposition will have the form $e^{\alpha c_i^\dagger c_{i'}} = 1 + \alpha c_i^\dagger c_{i'}$.}
\begin{equation}\label{app:exponential_trick}
    e^{\alpha c_i^\dagger c_i} = 
    1 + (e^\alpha -1) c_i^\dagger c_i,
    \qquad 
    c_i^\dagger c_i =  \frac{e^{\alpha c_i^\dagger c_j} - 1}{e^\alpha - 1}.
\end{equation}
With this, we can write 
\begin{equation}\label{app_tmp1}
    \tr \big\{c_i^\dagger c_i e^{\mathcal{X}} \big\} =
    \frac{1}{e^{\alpha} - 1} \Bigg[ \tr \big(e^{\alpha c_i^\dagger c_i} e^{\mathcal{X}} \big) - \tr\big(e^{\mathcal{X}} e^{\mathcal{Y}}\big)\Bigg].
\end{equation}
Eq.~\eqref{tr_det} is now applicable to each term individually. 
Here I have assumed only a single exponential $e^{\mathcal{X}}$. But since Eq.~\eqref{tr_det}  holds for an arbitrary number of quadratic forms, the results can be readily extended. 
Of course, the result must be independent of $\alpha$, so this constant must eventually factor out. 
In the first term of~\eqref{app_tmp1}, the quantity  $e^{\alpha c_i^\dagger c_i}$ as a quadratic form, with a matrix $R_{ii} = |i\rangle\langle i|$; i.e.,  with all elements being zero except the entry $(i,i)$. 
Here we also introduced for convenience the notation $|i\rangle$ to represent single particle kets (from a basis of $L$ elements). 
Thus we can write 
\begin{equation}
    \tr \big\{c_i^\dagger c_i e^{\mathcal{X}} \big\}
    = \frac{1}{e^{\alpha} - 1} \Big[ 
    \det (1 + e^{\alpha R_{ii}} e^{X} ) - \det (1 + e^{X})\Big].
\end{equation}
However, we also have that $e^{\alpha R_{ii}} = 1 + (e^\alpha - 1) |i\rangle\langle i|$, so the first term is written as 
\begin{equation}
    \det (1 + e^{\alpha Z_{ii}} e^{X} ) = \det\Big[1 + e^X + (e^\alpha - 1) |i\rangle\langle i| e^X\Big]. 
\end{equation}
Next we use the Sylvester determinant identity, which states that 
\begin{equation}\label{sylvester}
    \det(A + |\psi\rangle\langle \phi|) = \det(A) (1+\langle \phi|A^{-1} |\psi\rangle).
\end{equation}
This yields 
\begin{equation}
    \det (1 + e^{\alpha Z_{ii}} e^{X} ) =\det(1 + e^X) \Big[ 1 + (e^\alpha - 1) \langle i | e^X (1+e^X)^{-1} |i\rangle \Big].
\end{equation}
Plugging this in Eq.~\eqref{app_tmp1} finally leads to a cancelation of the factor $e^\alpha - 1$, as expected. The only thing left is 
\begin{equation}
    \tr \big\{c_i^\dagger c_i e^{\mathcal{X}} \big\} = \det(1+e^X) \langle i| e^X (1+e^X)^{-1} |i\rangle.
\end{equation}
This is almost Eq.~\eqref{tr_det_cdc}. 
To finish, we  extend it to multiple matrices, $X,Y,Z$, leading to 
\begin{equation}
    \tr \big\{c_i^\dagger c_i e^{\mathcal{X}}e^{\mathcal{Y}}e^{\mathcal{Z}} \big\} = \det(1+e^Xe^Y e^Z) \langle i| e^X e^Y e^Z (1+e^X e^Y e^Z)^{-1} |i\rangle.
\end{equation}
The form shown in Eq.~\eqref{tr_det_cdc} is finally obtained by writing, e.g., 
$e^X (1+e^X)^{-1} = (e^{-X} + 1)^{-1}$.

\subsection{Proof of Eq.~\eqref{tr_det_2_cs}}

Next, we turn to Eq.~\eqref{tr_det_2_cs}, which is harder. 
We again use the factorization in~\eqref{app:exponential_trick} to write 
\begin{IEEEeqnarray}{rCl}
\label{app:hard_tmp1}
\tr \Big\{ c_i^\dagger c_{i} e^{\mathcal{X}} c_j^\dagger c_{j} e^{\mathcal{Y}}e^{\mathcal{Z}} \Big\} &=& 
\frac{1}{e^\alpha -1} \Bigg\{
\tr \Big[ c_i^\dagger c_i e^{\mathcal{X}} e^{\alpha c_j^\dagger c_j} e^\mathcal{Y} e^\mathcal{Z} \Big] - \tr\Big[ c_i^\dagger c_i e^{\mathcal{X}} e^\mathcal{Y} e^\mathcal{Z}\Big] \Bigg\}.
\end{IEEEeqnarray}
Both terms can now be computed from Eq.~\eqref{tr_det_cdc}. 
The last is in fact \emph{exactly} Eq.~\eqref{tr_det_cdc}. 
For simplicity, we are going to define 
\begin{equation}\label{DT}
    \mathbb{D} = \det (1 + e^X e^Y e^Z), \qquad
    \mathcal{T} = ( e^{-Z} e^{-Y} e^{-X} + 1)^{-1} = e^X e^Y e^Z (1 + e^X e^Y e^Z)^{-1}.
\end{equation}
Then the last term in~\eqref{app:hard_tmp1} becomes 
\begin{equation}\label{app:hard_tmp6}
     \tr\Big[ c_i^\dagger c_i e^{\mathcal{X}} e^\mathcal{Y} e^\mathcal{Z}\Big] = 
     \mathbb{D} \mathcal{T}_{ii}.
\end{equation}
Conversely, the first term reads
\begin{equation}\label{app:hard_tmp2}
    \tr \Big[ c_i^\dagger c_i e^{\mathcal{X}} e^{\alpha c_j^\dagger c_j} e^\mathcal{Y} e^\mathcal{Z} \Big]  = \det(1 + e^X e^{\alpha R_{jj}} e^Y e^Z )~
    \Big[e^{-Z} e^{-Y} e^{-\alpha R_{jj}} e^{-X} + 1\Big]_{ii}^{-1}.
\end{equation}
This formula still requires some working. 
We again write $e^{\alpha R_{jj}} = 1 + (e^\alpha -1) |j\rangle\langle j|$. 
Using Sylvester's identity~\eqref{sylvester}, the part associated to the determinant can be written as 
\begin{IEEEeqnarray}{rCl}
\nonumber
\det(1 + e^X e^{\alpha R_{jj}} e^Y e^Z )
&=& 
\det \Big( 1 + e^X e^Y e^Z   + (e^\alpha -1)  e^X |j\rangle\langle j| e^Y e^Z \Big)
\\[0.2cm]
&=& \det (1 + e^X e^Y e^Z) \Big\{ 1 + (e^\alpha-1) \langle j | e^Y e^Z (1+e^X e^Y e^Z)^{-1} e^X |j \rangle \Big\}
\nonumber 
\\[0.2cm]
&=& \mathbb{D} \Big[ 1 + (e^\alpha-1) 
\big(
e^{-X} \mathcal{T} e^X
\big)_{jj} \Big],
\label{app:hard_tmp3}
\end{IEEEeqnarray}
where Eq.~\eqref{DT} was used in the last line.

To treat the second term in Eq.~\eqref{app:hard_tmp2}, we first write it as 
\begin{equation}
    \Big[e^{-Z} e^{-Y} e^{-\alpha R_{jj}} e^{-X} + 1\Big]^{-1} = 
    \Big[ 1 + e^{-Z} e^{-Y} e^{-X} + (e^{-\alpha}-1) e^{-Z} e^{-Y} |j\rangle\langle j| e^{-X} \Big]^{-1},
\end{equation}
and then use the Sherman-Morisson formula, which states that 
\begin{equation}\label{sherman_morisson}
    (A + |\psi\rangle\langle \phi|)^{-1} = A^{-1} - \frac{A^{-1} |\psi\rangle\langle \phi| A^{-1}}{1 + \langle \phi | A^{-1} |\psi\rangle}.
\end{equation}
In our case  $A = e^{-Z} e^{-Y} e^{-X} + 1$ so $A^{-1} \equiv \mathcal{T}$ [Eq.~\eqref{DT}]. 
As a result, we get that the $i,i$ element of this will be 
\begin{equation}\label{app:hard_tmp4}
    \Big[e^{-Z} e^{-Y} e^{-\alpha R_{jj}} e^{-X} + 1\Big]^{-1}_{ii} = 
    \mathcal{T}_{ii} -(e^{-\alpha} -1)\frac{(\mathcal{T} e^{-Z} e^{-Y})_{ij} (e^{-X} \mathcal{T})_{ji}}{1 + (e^{-\alpha} - 1) \big( e^{-X} \mathcal{T} e^{-Z} e^{-Y} \big)_{jj}}. 
\end{equation}

Inserting Eqs.~\eqref{app:hard_tmp3} and~\eqref{app:hard_tmp4} in Eq.~\eqref{app:hard_tmp2},
leads to 
\begin{IEEEeqnarray*}{rCl}
    \tr \Big[ c_i^\dagger c_i e^{\mathcal{X}} e^{\alpha c_j^\dagger c_j} e^\mathcal{Y} e^\mathcal{Z} \Big]  &=&
    \mathbb{D}
    \Big[ 1 + (e^\alpha-1) 
    \big(
    e^{-X} \mathcal{T} e^X
    \big)_{jj} \Big]
    \Big[ 
    \mathcal{T}_{ii} -(e^{-\alpha} -1)\frac{(\mathcal{T} e^{-Z} e^{-Y})_{ij} (e^{-X} \mathcal{T})_{ji}}{1 + (e^{-\alpha} - 1) \big( e^{-X} \mathcal{T} e^{-Z} e^{-Y} \big)_{jj}}\Big]
    \\[0.2cm]
    &=& \mathbb{D} \Bigg[ \mathcal{T}_{ii} + (e^\alpha-1) (e^{-X} \mathcal{T} e^X)_{jj} \mathcal{T}_{ii} - 
    (e^{-\alpha}-1) 
    \frac{(\mathcal{T} e^{-Z} e^{-Y})_{ij} (e^{-X} \mathcal{T})_{ji}}{1 + (e^{-\alpha} - 1) \big( e^{-X} \mathcal{T} e^{-Z} e^{-Y} \big)_{jj}}
    \\[0.2cm]
    &&- (e^\alpha-1)(e^{-\alpha} - 1)
    \frac{  (e^{-X} \mathcal{T} e^X)_{jj}(\mathcal{T} e^{-Z} e^{-Y})_{ij} (e^{-X} \mathcal{T})_{ji}}{1 + (e^{-\alpha} - 1) \big[ e^{-X} \mathcal{T} e^{-Z} e^{-Y} \big]_{jj}}
    \Bigg]
\end{IEEEeqnarray*}
Finally, we insert this in Eq.~\eqref{app:hard_tmp1}.
In light of Eq.~\eqref{app:hard_tmp6}, this simply means we cancel out the term $\mathbb{D} \mathcal{T}_{ii}$. Hence, we are only left with 
\begin{IEEEeqnarray*}{rCl}
    \tr \Big\{ c_i^\dagger c_{i} e^{\mathcal{X}} c_j^\dagger c_{j} e^{\mathcal{Y}}e^{\mathcal{Z}} \Big\} &=& \mathbb{D}\Bigg\{
    (e^{-X} \mathcal{T} e^X)_{jj} \mathcal{T}_{ii}
    +
     (\mathcal{T} e^{-Z} e^{-Y})_{ij} (e^{-X} \mathcal{T})_{ji}
     \frac{ (e^\alpha - 1)(e^{-X} \mathcal{T} e^{X})_{jj}+1}{ e^\alpha \big[ 1 - (e^{-X} \mathcal{T} e^{-Z} e^{-Y})_{jj}\big] + (e^{-X} \mathcal{T} e^{-Z} e^{-Y})_{jj}} \Bigg\}
\end{IEEEeqnarray*}
Using the structure of $\mathcal{T}$ in Eq.~\eqref{DT}, one may verify that the matrix appearing in the denominator is actually related to the matrix $e^{-X} \mathcal{T} e^X$ according to 
\begin{equation}
    e^{-X} \mathcal{T} e^{-Z} e^{-Y} = 1 - e^{-X} \mathcal{T} e^X.
\end{equation}
This allows for the expression to be simplified, finally leading to a cancelation of the factor of $e^\alpha$ (as it must, since $\alpha$ is arbitrary). 
As a result, we are left only with 
\begin{equation}
    \tr \Big\{ c_i^\dagger c_{i} e^{\mathcal{X}} c_j^\dagger c_{j} e^{\mathcal{Y}}e^{\mathcal{Z}} \Big\}
    = \mathbb{D}\Bigg\{
    (e^{-X} \mathcal{T} e^X)_{jj} \mathcal{T}_{ii}
    +
     (\mathcal{T} e^{-Z} e^{-Y})_{ij} (e^{-X} \mathcal{T})_{ji} \Bigg\}.
\end{equation}
The formula in the case when $i' \neq i$ and $j' \neq j$ is similar, and reads
\begin{equation}
    \tr \Big\{ c_i^\dagger c_{i'} e^{\mathcal{X}} c_j^\dagger c_{j'} e^{\mathcal{Y}}e^{\mathcal{Z}} \Big\}
    = \mathbb{D}\Bigg\{
    (e^{-X} \mathcal{T} e^X)_{j'j} \mathcal{T}_{i'i}
    +
     (\mathcal{T} e^{-Z} e^{-Y})_{i'j} (e^{-X} \mathcal{T})_{j'i} \Bigg\}.
\end{equation}

\end{widetext}
\bibliography{library}

\begin{thebibliography}{81}%
\makeatletter
\providecommand \@ifxundefined [1]{%
 \@ifx{#1\undefined}
}%
\providecommand \@ifnum [1]{%
 \ifnum #1\expandafter \@firstoftwo
 \else \expandafter \@secondoftwo
 \fi
}%
\providecommand \@ifx [1]{%
 \ifx #1\expandafter \@firstoftwo
 \else \expandafter \@secondoftwo
 \fi
}%
\providecommand \natexlab [1]{#1}%
\providecommand \enquote  [1]{``#1''}%
\providecommand \bibnamefont  [1]{#1}%
\providecommand \bibfnamefont [1]{#1}%
\providecommand \citenamefont [1]{#1}%
\providecommand \href@noop [0]{\@secondoftwo}%
\providecommand \href [0]{\begingroup \@sanitize@url \@href}%
\providecommand \@href[1]{\@@startlink{#1}\@@href}%
\providecommand \@@href[1]{\endgroup#1\@@endlink}%
\providecommand \@sanitize@url [0]{\catcode `\\12\catcode `\$12\catcode
  `\&12\catcode `\#12\catcode `\^12\catcode `\_12\catcode `\%12\relax}%
\providecommand \@@startlink[1]{}%
\providecommand \@@endlink[0]{}%
\providecommand \url  [0]{\begingroup\@sanitize@url \@url }%
\providecommand \@url [1]{\endgroup\@href {#1}{\urlprefix }}%
\providecommand \urlprefix  [0]{URL }%
\providecommand \Eprint [0]{\href }%
\providecommand \doibase [0]{http://dx.doi.org/}%
\providecommand \selectlanguage [0]{\@gobble}%
\providecommand \bibinfo  [0]{\@secondoftwo}%
\providecommand \bibfield  [0]{\@secondoftwo}%
\providecommand \translation [1]{[#1]}%
\providecommand \BibitemOpen [0]{}%
\providecommand \bibitemStop [0]{}%
\providecommand \bibitemNoStop [0]{.\EOS\space}%
\providecommand \EOS [0]{\spacefactor3000\relax}%
\providecommand \BibitemShut  [1]{\csname bibitem#1\endcsname}%
\let\auto@bib@innerbib\@empty
\bibitem [{\citenamefont {Bertini}\ \emph {et~al.}(2021)\citenamefont
  {Bertini}, \citenamefont {Heidrich-Meisner}, \citenamefont {Karrasch},
  \citenamefont {Prosen}, \citenamefont {Steinigeweg},\ and\ \citenamefont
  {{\v{Z}}nidari{\v{c}}}}]{Bertini2020}%
  \BibitemOpen
  \bibfield  {author} {\bibinfo {author} {\bibfnamefont {B.}~\bibnamefont
  {Bertini}}, \bibinfo {author} {\bibfnamefont {F.}~\bibnamefont
  {Heidrich-Meisner}}, \bibinfo {author} {\bibfnamefont {C.}~\bibnamefont
  {Karrasch}}, \bibinfo {author} {\bibfnamefont {T.}~\bibnamefont {Prosen}},
  \bibinfo {author} {\bibfnamefont {R.}~\bibnamefont {Steinigeweg}}, \ and\
  \bibinfo {author} {\bibfnamefont {M.}~\bibnamefont {{\v{Z}}nidari{\v{c}}}},\
  }\href {\doibase 10.1103/RevModPhys.93.025003} {\bibfield  {journal}
  {\bibinfo  {journal} {Reviews of Modern Physics}\ }\textbf {\bibinfo {volume}
  {93}},\ \bibinfo {pages} {025003} (\bibinfo {year} {2021})}\BibitemShut
  {NoStop}%
\bibitem [{\citenamefont {{\v{Z}}nidari{\v{c}}}(2011)}]{Znidaric2011}%
  \BibitemOpen
  \bibfield  {author} {\bibinfo {author} {\bibfnamefont {M.}~\bibnamefont
  {{\v{Z}}nidari{\v{c}}}},\ }\href {\doibase 10.1103/PhysRevLett.106.220601}
  {\bibfield  {journal} {\bibinfo  {journal} {Physical Review Letters}\
  }\textbf {\bibinfo {volume} {106}},\ \bibinfo {pages} {220601} (\bibinfo
  {year} {2011})}\BibitemShut {NoStop}%
\bibitem [{\citenamefont {Landi}\ and\ \citenamefont
  {Karevski}(2015)}]{Landi2015a}%
  \BibitemOpen
  \bibfield  {author} {\bibinfo {author} {\bibfnamefont {G.~T.}\ \bibnamefont
  {Landi}}\ and\ \bibinfo {author} {\bibfnamefont {D.}~\bibnamefont
  {Karevski}},\ }\href {\doibase 10.1103/PhysRevB.91.174422} {\bibfield
  {journal} {\bibinfo  {journal} {Physical Review B}\ }\textbf {\bibinfo
  {volume} {91}},\ \bibinfo {pages} {174422} (\bibinfo {year}
  {2015})}\BibitemShut {NoStop}%
\bibitem [{\citenamefont {Gopalakrishnan}\ and\ \citenamefont
  {Vasseur}(2019)}]{Gopalakrishnan2019}%
  \BibitemOpen
  \bibfield  {author} {\bibinfo {author} {\bibfnamefont {S.}~\bibnamefont
  {Gopalakrishnan}}\ and\ \bibinfo {author} {\bibfnamefont {R.}~\bibnamefont
  {Vasseur}},\ }\href {\doibase 10.1103/PhysRevLett.122.127202} {\bibfield
  {journal} {\bibinfo  {journal} {Physical Review Letters}\ }\textbf {\bibinfo
  {volume} {122}},\ \bibinfo {pages} {127202} (\bibinfo {year}
  {2019})}\BibitemShut {NoStop}%
\bibitem [{\citenamefont {Bulchandani}\ and\ \citenamefont
  {Karrasch}(2019)}]{Bulchandani2019}%
  \BibitemOpen
  \bibfield  {author} {\bibinfo {author} {\bibfnamefont {V.~B.}\ \bibnamefont
  {Bulchandani}}\ and\ \bibinfo {author} {\bibfnamefont {C.}~\bibnamefont
  {Karrasch}},\ }\href {\doibase 10.1103/PhysRevB.99.121410} {\bibfield
  {journal} {\bibinfo  {journal} {Physical Review B}\ }\textbf {\bibinfo
  {volume} {99}},\ \bibinfo {pages} {1} (\bibinfo {year} {2019})}\BibitemShut
  {NoStop}%
\bibitem [{\citenamefont {Ilievski}\ \emph {et~al.}(2018)\citenamefont
  {Ilievski}, \citenamefont {De~Nardis}, \citenamefont {Medenjak},\ and\
  \citenamefont {Prosen}}]{Ilievski2018}%
  \BibitemOpen
  \bibfield  {author} {\bibinfo {author} {\bibfnamefont {E.}~\bibnamefont
  {Ilievski}}, \bibinfo {author} {\bibfnamefont {J.}~\bibnamefont {De~Nardis}},
  \bibinfo {author} {\bibfnamefont {M.}~\bibnamefont {Medenjak}}, \ and\
  \bibinfo {author} {\bibfnamefont {T.}~\bibnamefont {Prosen}},\ }\href
  {\doibase 10.1103/PhysRevLett.121.230602} {\bibfield  {journal} {\bibinfo
  {journal} {Physical Review Letters}\ }\textbf {\bibinfo {volume} {121}},\
  \bibinfo {pages} {230602} (\bibinfo {year} {2018})}\BibitemShut {NoStop}%
\bibitem [{\citenamefont {Viciani}\ \emph {et~al.}(2015)\citenamefont
  {Viciani}, \citenamefont {Lima}, \citenamefont {Bellini},\ and\ \citenamefont
  {Caruso}}]{Viciani2015}%
  \BibitemOpen
  \bibfield  {author} {\bibinfo {author} {\bibfnamefont {S.}~\bibnamefont
  {Viciani}}, \bibinfo {author} {\bibfnamefont {M.}~\bibnamefont {Lima}},
  \bibinfo {author} {\bibfnamefont {M.}~\bibnamefont {Bellini}}, \ and\
  \bibinfo {author} {\bibfnamefont {F.}~\bibnamefont {Caruso}},\ }\href
  {\doibase 10.1103/PhysRevLett.115.083601} {\bibfield  {journal} {\bibinfo
  {journal} {Physical Review Letters}\ }\textbf {\bibinfo {volume} {115}},\
  \bibinfo {pages} {083601} (\bibinfo {year} {2015})}\BibitemShut {NoStop}%
\bibitem [{\citenamefont {Plenio}\ and\ \citenamefont
  {Huelga}(2008)}]{Plenio2008}%
  \BibitemOpen
  \bibfield  {author} {\bibinfo {author} {\bibfnamefont {M.~B.}\ \bibnamefont
  {Plenio}}\ and\ \bibinfo {author} {\bibfnamefont {S.~F.}\ \bibnamefont
  {Huelga}},\ }\href {\doibase 10.1088/1367-2630/10/11/113019} {\bibfield
  {journal} {\bibinfo  {journal} {New Journal of Physics}\ }\textbf {\bibinfo
  {volume} {10}},\ \bibinfo {pages} {0} (\bibinfo {year} {2008})}\BibitemShut
  {NoStop}%
\bibitem [{\citenamefont {Biggerstaff}\ \emph {et~al.}(2016)\citenamefont
  {Biggerstaff}, \citenamefont {Heilmann}, \citenamefont {Zecevik},
  \citenamefont {Gr{\"{a}}fe}, \citenamefont {Broome}, \citenamefont
  {Fedrizzi}, \citenamefont {Nolte}, \citenamefont {Szameit}, \citenamefont
  {White},\ and\ \citenamefont {Kassal}}]{Biggerstaff2016}%
  \BibitemOpen
  \bibfield  {author} {\bibinfo {author} {\bibfnamefont {D.~N.}\ \bibnamefont
  {Biggerstaff}}, \bibinfo {author} {\bibfnamefont {R.}~\bibnamefont
  {Heilmann}}, \bibinfo {author} {\bibfnamefont {A.~A.}\ \bibnamefont
  {Zecevik}}, \bibinfo {author} {\bibfnamefont {M.}~\bibnamefont
  {Gr{\"{a}}fe}}, \bibinfo {author} {\bibfnamefont {M.~A.}\ \bibnamefont
  {Broome}}, \bibinfo {author} {\bibfnamefont {A.}~\bibnamefont {Fedrizzi}},
  \bibinfo {author} {\bibfnamefont {S.}~\bibnamefont {Nolte}}, \bibinfo
  {author} {\bibfnamefont {A.}~\bibnamefont {Szameit}}, \bibinfo {author}
  {\bibfnamefont {A.~G.}\ \bibnamefont {White}}, \ and\ \bibinfo {author}
  {\bibfnamefont {I.}~\bibnamefont {Kassal}},\ }\href {\doibase
  10.1038/ncomms11282} {\bibfield  {journal} {\bibinfo  {journal} {Nature
  Communications}\ }\textbf {\bibinfo {volume} {7}},\ \bibinfo {pages} {1}
  (\bibinfo {year} {2016})}\BibitemShut {NoStop}%
\bibitem [{\citenamefont {Maier}\ \emph {et~al.}(2019)\citenamefont {Maier},
  \citenamefont {Brydges}, \citenamefont {Jurcevic}, \citenamefont {Trautmann},
  \citenamefont {Hempel}, \citenamefont {Lanyon}, \citenamefont {Hauke},
  \citenamefont {Blatt},\ and\ \citenamefont {Roos}}]{Maier2019}%
  \BibitemOpen
  \bibfield  {author} {\bibinfo {author} {\bibfnamefont {C.}~\bibnamefont
  {Maier}}, \bibinfo {author} {\bibfnamefont {T.}~\bibnamefont {Brydges}},
  \bibinfo {author} {\bibfnamefont {P.}~\bibnamefont {Jurcevic}}, \bibinfo
  {author} {\bibfnamefont {N.}~\bibnamefont {Trautmann}}, \bibinfo {author}
  {\bibfnamefont {C.}~\bibnamefont {Hempel}}, \bibinfo {author} {\bibfnamefont
  {B.~P.}\ \bibnamefont {Lanyon}}, \bibinfo {author} {\bibfnamefont
  {P.}~\bibnamefont {Hauke}}, \bibinfo {author} {\bibfnamefont
  {R.}~\bibnamefont {Blatt}}, \ and\ \bibinfo {author} {\bibfnamefont {C.~F.}\
  \bibnamefont {Roos}},\ }\href {\doibase 10.1103/PhysRevLett.122.050501}
  {\bibfield  {journal} {\bibinfo  {journal} {Physical Review Letters}\
  }\textbf {\bibinfo {volume} {122}},\ \bibinfo {pages} {050501} (\bibinfo
  {year} {2019})}\BibitemShut {NoStop}%
\bibitem [{\citenamefont {De~Le{\'{o}}n-Montiel}\ \emph
  {et~al.}(2015)\citenamefont {De~Le{\'{o}}n-Montiel}, \citenamefont
  {Quiroz-Ju{\'{a}}rez}, \citenamefont {Quintero-Torres}, \citenamefont
  {Dom{\'{i}}nguez-Ju{\'{a}}rez}, \citenamefont {Moya-Cessa}, \citenamefont
  {Torres},\ and\ \citenamefont {Arag{\'{o}}n}}]{DeLeon-Montiel2015}%
  \BibitemOpen
  \bibfield  {author} {\bibinfo {author} {\bibfnamefont {J.~R.}\ \bibnamefont
  {De~Le{\'{o}}n-Montiel}}, \bibinfo {author} {\bibfnamefont {M.~A.}\
  \bibnamefont {Quiroz-Ju{\'{a}}rez}}, \bibinfo {author} {\bibfnamefont
  {R.}~\bibnamefont {Quintero-Torres}}, \bibinfo {author} {\bibfnamefont
  {J.~L.}\ \bibnamefont {Dom{\'{i}}nguez-Ju{\'{a}}rez}}, \bibinfo {author}
  {\bibfnamefont {H.~M.}\ \bibnamefont {Moya-Cessa}}, \bibinfo {author}
  {\bibfnamefont {J.~P.}\ \bibnamefont {Torres}}, \ and\ \bibinfo {author}
  {\bibfnamefont {J.~L.}\ \bibnamefont {Arag{\'{o}}n}},\ }\href {\doibase
  10.1038/srep17339} {\bibfield  {journal} {\bibinfo  {journal} {Scientific
  Reports}\ }\textbf {\bibinfo {volume} {5}},\ \bibinfo {pages} {1} (\bibinfo
  {year} {2015})}\BibitemShut {NoStop}%
\bibitem [{\citenamefont {Dwiputra}\ and\ \citenamefont
  {Zen}(2020)}]{Dwiputra2020}%
  \BibitemOpen
  \bibfield  {author} {\bibinfo {author} {\bibfnamefont {D.}~\bibnamefont
  {Dwiputra}}\ and\ \bibinfo {author} {\bibfnamefont {F.~P.}\ \bibnamefont
  {Zen}},\ }\href@noop {} {\bibfield  {journal} {\bibinfo  {journal} {arXiv}\
  ,\ \bibinfo {pages} {1}} (\bibinfo {year} {2020})}\BibitemShut {NoStop}%
\bibitem [{\citenamefont {Benenti}\ \emph {et~al.}(2017)\citenamefont
  {Benenti}, \citenamefont {Casati}, \citenamefont {Saito},\ and\ \citenamefont
  {Whitney}}]{Benenti2017a}%
  \BibitemOpen
  \bibfield  {author} {\bibinfo {author} {\bibfnamefont {G.}~\bibnamefont
  {Benenti}}, \bibinfo {author} {\bibfnamefont {G.}~\bibnamefont {Casati}},
  \bibinfo {author} {\bibfnamefont {K.}~\bibnamefont {Saito}}, \ and\ \bibinfo
  {author} {\bibfnamefont {R.~S.}\ \bibnamefont {Whitney}},\ }\href {\doibase
  10.1016/j.physrep.2017.05.008} {\bibfield  {journal} {\bibinfo  {journal}
  {Physics Reports}\ }\textbf {\bibinfo {volume} {694}},\ \bibinfo {pages} {1}
  (\bibinfo {year} {2017})}\BibitemShut {NoStop}%
\bibitem [{\citenamefont {Mahan}\ and\ \citenamefont {Sofo}(1996)}]{Mahan1996}%
  \BibitemOpen
  \bibfield  {author} {\bibinfo {author} {\bibfnamefont {G.~D.}\ \bibnamefont
  {Mahan}}\ and\ \bibinfo {author} {\bibfnamefont {J.~O.}\ \bibnamefont
  {Sofo}},\ }\href {\doibase 10.1073/pnas.93.15.7436} {\bibfield  {journal}
  {\bibinfo  {journal} {Proceedings of the National Academy of Sciences of the
  United States of America}\ }\textbf {\bibinfo {volume} {93}},\ \bibinfo
  {pages} {7436} (\bibinfo {year} {1996})}\BibitemShut {NoStop}%
\bibitem [{\citenamefont {Yamamoto}\ and\ \citenamefont
  {Hatano}(2015)}]{Yamamoto2015a}%
  \BibitemOpen
  \bibfield  {author} {\bibinfo {author} {\bibfnamefont {K.}~\bibnamefont
  {Yamamoto}}\ and\ \bibinfo {author} {\bibfnamefont {N.}~\bibnamefont
  {Hatano}},\ }\href {\doibase 10.1103/PhysRevE.92.042165} {\bibfield
  {journal} {\bibinfo  {journal} {Physical Review E}\ }\textbf {\bibinfo
  {volume} {92}},\ \bibinfo {pages} {042165} (\bibinfo {year}
  {2015})}\BibitemShut {NoStop}%
\bibitem [{\citenamefont {Dubi}\ and\ \citenamefont
  {Di~Ventra}(2011)}]{Dubi2011}%
  \BibitemOpen
  \bibfield  {author} {\bibinfo {author} {\bibfnamefont {Y.}~\bibnamefont
  {Dubi}}\ and\ \bibinfo {author} {\bibfnamefont {M.}~\bibnamefont
  {Di~Ventra}},\ }\href {\doibase 10.1103/RevModPhys.83.131} {\bibfield
  {journal} {\bibinfo  {journal} {Reviews of Modern Physics}\ }\textbf
  {\bibinfo {volume} {83}},\ \bibinfo {pages} {131} (\bibinfo {year}
  {2011})}\BibitemShut {NoStop}%
\bibitem [{\citenamefont {Whitney}(2014)}]{Whitney2014}%
  \BibitemOpen
  \bibfield  {author} {\bibinfo {author} {\bibfnamefont {R.~S.}\ \bibnamefont
  {Whitney}},\ }\href {\doibase 10.1103/PhysRevLett.112.130601} {\bibfield
  {journal} {\bibinfo  {journal} {Physical Review Letters}\ }\textbf {\bibinfo
  {volume} {112}},\ \bibinfo {pages} {130601} (\bibinfo {year}
  {2014})}\BibitemShut {NoStop}%
\bibitem [{\citenamefont {Li}\ \emph {et~al.}(2012)\citenamefont {Li},
  \citenamefont {Ren}, \citenamefont {Wang}, \citenamefont {Zhang},
  \citenamefont {H{\"{a}}nggi},\ and\ \citenamefont {Li}}]{Li2012}%
  \BibitemOpen
  \bibfield  {author} {\bibinfo {author} {\bibfnamefont {N.}~\bibnamefont
  {Li}}, \bibinfo {author} {\bibfnamefont {J.}~\bibnamefont {Ren}}, \bibinfo
  {author} {\bibfnamefont {L.}~\bibnamefont {Wang}}, \bibinfo {author}
  {\bibfnamefont {G.}~\bibnamefont {Zhang}}, \bibinfo {author} {\bibfnamefont
  {P.}~\bibnamefont {H{\"{a}}nggi}}, \ and\ \bibinfo {author} {\bibfnamefont
  {B.}~\bibnamefont {Li}},\ }\href {\doibase 10.1103/RevModPhys.84.1045}
  {\bibfield  {journal} {\bibinfo  {journal} {Reviews of Modern Physics}\
  }\textbf {\bibinfo {volume} {84}},\ \bibinfo {pages} {1045} (\bibinfo {year}
  {2012})}\BibitemShut {NoStop}%
\bibitem [{\citenamefont {Pereira}\ and\ \citenamefont
  {{\'{A}}vila}(2013)}]{Pereira2013a}%
  \BibitemOpen
  \bibfield  {author} {\bibinfo {author} {\bibfnamefont {E.}~\bibnamefont
  {Pereira}}\ and\ \bibinfo {author} {\bibfnamefont {R.~R.}\ \bibnamefont
  {{\'{A}}vila}},\ }\href {\doibase 10.1103/PhysRevE.88.032139} {\bibfield
  {journal} {\bibinfo  {journal} {Physical Review E}\ }\textbf {\bibinfo
  {volume} {88}},\ \bibinfo {pages} {032139} (\bibinfo {year}
  {2013})}\BibitemShut {NoStop}%
\bibitem [{\citenamefont {Werlang}\ \emph {et~al.}(2014)\citenamefont
  {Werlang}, \citenamefont {Marchiori}, \citenamefont {Cornelio},\ and\
  \citenamefont {Valente}}]{Werlang2014}%
  \BibitemOpen
  \bibfield  {author} {\bibinfo {author} {\bibfnamefont {T.}~\bibnamefont
  {Werlang}}, \bibinfo {author} {\bibfnamefont {M.~A.}\ \bibnamefont
  {Marchiori}}, \bibinfo {author} {\bibfnamefont {M.~F.}\ \bibnamefont
  {Cornelio}}, \ and\ \bibinfo {author} {\bibfnamefont {D.}~\bibnamefont
  {Valente}},\ }\href@noop {} {\bibfield  {journal} {\bibinfo  {journal}
  {arXiv}\ } (\bibinfo {year} {2014})}\BibitemShut {NoStop}%
\bibitem [{\citenamefont {{\'{A}}vila}\ and\ \citenamefont
  {Pereira}(2013)}]{Avila2013}%
  \BibitemOpen
  \bibfield  {author} {\bibinfo {author} {\bibfnamefont {R.~R.}\ \bibnamefont
  {{\'{A}}vila}}\ and\ \bibinfo {author} {\bibfnamefont {E.}~\bibnamefont
  {Pereira}},\ }\href {\doibase 10.1088/1751-8113/46/5/055002} {\bibfield
  {journal} {\bibinfo  {journal} {Journal of Physics A: Mathematical and
  Theoretical}\ }\textbf {\bibinfo {volume} {46}},\ \bibinfo {pages} {055002}
  (\bibinfo {year} {2013})}\BibitemShut {NoStop}%
\bibitem [{\citenamefont {Schuab}\ \emph {et~al.}(2016)\citenamefont {Schuab},
  \citenamefont {Pereira},\ and\ \citenamefont {Landi}}]{Schuab2016a}%
  \BibitemOpen
  \bibfield  {author} {\bibinfo {author} {\bibfnamefont {L.}~\bibnamefont
  {Schuab}}, \bibinfo {author} {\bibfnamefont {E.}~\bibnamefont {Pereira}}, \
  and\ \bibinfo {author} {\bibfnamefont {G.~T.}\ \bibnamefont {Landi}},\ }\href
  {\doibase 10.1103/PhysRevE.94.042122} {\bibfield  {journal} {\bibinfo
  {journal} {Physical Review E}\ }\textbf {\bibinfo {volume} {94}},\ \bibinfo
  {pages} {042122} (\bibinfo {year} {2016})}\BibitemShut {NoStop}%
\bibitem [{\citenamefont {Balachandran}\ \emph {et~al.}(2018)\citenamefont
  {Balachandran}, \citenamefont {Benenti}, \citenamefont {Pereira},
  \citenamefont {Casati},\ and\ \citenamefont {Poletti}}]{Balachandran2018}%
  \BibitemOpen
  \bibfield  {author} {\bibinfo {author} {\bibfnamefont {V.}~\bibnamefont
  {Balachandran}}, \bibinfo {author} {\bibfnamefont {G.}~\bibnamefont
  {Benenti}}, \bibinfo {author} {\bibfnamefont {E.}~\bibnamefont {Pereira}},
  \bibinfo {author} {\bibfnamefont {G.}~\bibnamefont {Casati}}, \ and\ \bibinfo
  {author} {\bibfnamefont {D.}~\bibnamefont {Poletti}},\ }\href {\doibase
  10.1103/PhysRevLett.120.200603} {\bibfield  {journal} {\bibinfo  {journal}
  {Physical Review Letters}\ }\textbf {\bibinfo {volume} {120}},\ \bibinfo
  {pages} {200603} (\bibinfo {year} {2018})}\BibitemShut {NoStop}%
\bibitem [{\citenamefont {Pereira}(2010{\natexlab{a}})}]{Pereira2010b}%
  \BibitemOpen
  \bibfield  {author} {\bibinfo {author} {\bibfnamefont {E.}~\bibnamefont
  {Pereira}},\ }\href {\doibase 10.1016/j.physleta.2010.02.071} {\bibfield
  {journal} {\bibinfo  {journal} {Physics Letters A}\ }\textbf {\bibinfo
  {volume} {374}},\ \bibinfo {pages} {1933} (\bibinfo {year}
  {2010}{\natexlab{a}})}\BibitemShut {NoStop}%
\bibitem [{\citenamefont {Pereira}(2010{\natexlab{b}})}]{Pereira2010}%
  \BibitemOpen
  \bibfield  {author} {\bibinfo {author} {\bibfnamefont {E.}~\bibnamefont
  {Pereira}},\ }\href {\doibase 10.1103/PhysRevE.82.040101} {\bibfield
  {journal} {\bibinfo  {journal} {Physical Review E}\ }\textbf {\bibinfo
  {volume} {82}},\ \bibinfo {pages} {040101} (\bibinfo {year}
  {2010}{\natexlab{b}})}\BibitemShut {NoStop}%
\bibitem [{\citenamefont {Wang}\ and\ \citenamefont {Li}(2007)}]{Wang2007}%
  \BibitemOpen
  \bibfield  {author} {\bibinfo {author} {\bibfnamefont {L.}~\bibnamefont
  {Wang}}\ and\ \bibinfo {author} {\bibfnamefont {B.}~\bibnamefont {Li}},\
  }\href {\doibase 10.1103/PhysRevLett.99.177208} {\bibfield  {journal}
  {\bibinfo  {journal} {Physical Review Letters}\ }\textbf {\bibinfo {volume}
  {99}},\ \bibinfo {pages} {177208} (\bibinfo {year} {2007})}\BibitemShut
  {NoStop}%
\bibitem [{\citenamefont {Hu}\ \emph {et~al.}(2006)\citenamefont {Hu},
  \citenamefont {Yang},\ and\ \citenamefont {Zhang}}]{Hu2006}%
  \BibitemOpen
  \bibfield  {author} {\bibinfo {author} {\bibfnamefont {B.}~\bibnamefont
  {Hu}}, \bibinfo {author} {\bibfnamefont {L.}~\bibnamefont {Yang}}, \ and\
  \bibinfo {author} {\bibfnamefont {Y.}~\bibnamefont {Zhang}},\ }\href
  {\doibase 10.1103/PhysRevLett.97.124302} {\bibfield  {journal} {\bibinfo
  {journal} {Physical Review Letters}\ }\textbf {\bibinfo {volume} {97}},\
  \bibinfo {pages} {124302} (\bibinfo {year} {2006})}\BibitemShut {NoStop}%
\bibitem [{\citenamefont {Landi}\ \emph {et~al.}(2014)\citenamefont {Landi},
  \citenamefont {Novais}, \citenamefont {de~Oliveira},\ and\ \citenamefont
  {Karevski}}]{Landi2014b}%
  \BibitemOpen
  \bibfield  {author} {\bibinfo {author} {\bibfnamefont {G.~T.}\ \bibnamefont
  {Landi}}, \bibinfo {author} {\bibfnamefont {E.}~\bibnamefont {Novais}},
  \bibinfo {author} {\bibfnamefont {M.~J.}\ \bibnamefont {de~Oliveira}}, \ and\
  \bibinfo {author} {\bibfnamefont {D.}~\bibnamefont {Karevski}},\ }\href
  {\doibase 10.1103/PhysRevE.90.042142} {\bibfield  {journal} {\bibinfo
  {journal} {Physical Review E}\ }\textbf {\bibinfo {volume} {90}},\ \bibinfo
  {pages} {042142} (\bibinfo {year} {2014})}\BibitemShut {NoStop}%
\bibitem [{\citenamefont {Silva}\ \emph {et~al.}(2020)\citenamefont {Silva},
  \citenamefont {Landi}, \citenamefont {Drumond},\ and\ \citenamefont
  {Pereira}}]{Silva2020}%
  \BibitemOpen
  \bibfield  {author} {\bibinfo {author} {\bibfnamefont {S.~H.~S.}\
  \bibnamefont {Silva}}, \bibinfo {author} {\bibfnamefont {G.~T.}\ \bibnamefont
  {Landi}}, \bibinfo {author} {\bibfnamefont {R.~C.}\ \bibnamefont {Drumond}},
  \ and\ \bibinfo {author} {\bibfnamefont {E.}~\bibnamefont {Pereira}},\ }\href
  {\doibase 10.1103/PhysRevE.102.062146} {\bibfield  {journal} {\bibinfo
  {journal} {Physical Review E}\ }\textbf {\bibinfo {volume} {102}},\ \bibinfo
  {pages} {062146} (\bibinfo {year} {2020})}\BibitemShut {NoStop}%
\bibitem [{\citenamefont {Chioquetta}\ \emph {et~al.}(2021)\citenamefont
  {Chioquetta}, \citenamefont {Pereira}, \citenamefont {Landi},\ and\
  \citenamefont {Drumond}}]{Chioquetta2021}%
  \BibitemOpen
  \bibfield  {author} {\bibinfo {author} {\bibfnamefont {A.}~\bibnamefont
  {Chioquetta}}, \bibinfo {author} {\bibfnamefont {E.}~\bibnamefont {Pereira}},
  \bibinfo {author} {\bibfnamefont {G.~T.}\ \bibnamefont {Landi}}, \ and\
  \bibinfo {author} {\bibfnamefont {R.~C.}\ \bibnamefont {Drumond}},\ }\href
  {\doibase 10.1103/PhysRevE.103.032108} {\bibfield  {journal} {\bibinfo
  {journal} {Physical Review E}\ }\textbf {\bibinfo {volume} {103}},\ \bibinfo
  {pages} {032108} (\bibinfo {year} {2021})}\BibitemShut {NoStop}%
\bibitem [{\citenamefont {Datta}(1997)}]{Datta1997a}%
  \BibitemOpen
  \bibfield  {author} {\bibinfo {author} {\bibfnamefont {S.}~\bibnamefont
  {Datta}},\ }\href@noop {} {\emph {\bibinfo {title} {{Electronic Transport in
  Mesoscopic Systems}}}}\ (\bibinfo  {publisher} {Cambridge University Press},\
  \bibinfo {address} {Cambridge, UK},\ \bibinfo {year} {1997})\BibitemShut
  {NoStop}%
\bibitem [{\citenamefont {Landi}\ \emph {et~al.}(2021)\citenamefont {Landi},
  \citenamefont {Schaller},\ and\ \citenamefont {Poletti}}]{Landi2021}%
  \BibitemOpen
  \bibfield  {author} {\bibinfo {author} {\bibfnamefont {G.~T.}\ \bibnamefont
  {Landi}}, \bibinfo {author} {\bibfnamefont {G.}~\bibnamefont {Schaller}}, \
  and\ \bibinfo {author} {\bibfnamefont {D.}~\bibnamefont {Poletti}},\
  }\href@noop {} {\bibfield  {journal} {\bibinfo  {journal} {To appear in
  Review of Modern Physics}\ } (\bibinfo {year} {2021})}\BibitemShut {NoStop}%
\bibitem [{\citenamefont {Levitov}\ and\ \citenamefont
  {Lesovik}(1993)}]{Levitov1993}%
  \BibitemOpen
  \bibfield  {author} {\bibinfo {author} {\bibfnamefont {L.}~\bibnamefont
  {Levitov}}\ and\ \bibinfo {author} {\bibfnamefont {G.}~\bibnamefont
  {Lesovik}},\ }\href@noop {} {\bibfield  {journal} {\bibinfo  {journal} {JETP
  letters}\ }\textbf {\bibinfo {volume} {58}},\ \bibinfo {pages} {230}
  (\bibinfo {year} {1993})}\BibitemShut {NoStop}%
\bibitem [{\citenamefont {Esposito}\ \emph {et~al.}(2007)\citenamefont
  {Esposito}, \citenamefont {Harbola},\ and\ \citenamefont
  {Mukamel}}]{Esposito2007}%
  \BibitemOpen
  \bibfield  {author} {\bibinfo {author} {\bibfnamefont {M.}~\bibnamefont
  {Esposito}}, \bibinfo {author} {\bibfnamefont {U.}~\bibnamefont {Harbola}}, \
  and\ \bibinfo {author} {\bibfnamefont {S.}~\bibnamefont {Mukamel}},\ }\href
  {\doibase 10.1103/PhysRevB.75.155316} {\bibfield  {journal} {\bibinfo
  {journal} {Physical Review B}\ }\textbf {\bibinfo {volume} {75}},\ \bibinfo
  {pages} {155316} (\bibinfo {year} {2007})},\ \Eprint
  {http://arxiv.org/abs/0702376v1} {arXiv:0702376v1 [cond-mat]} \BibitemShut
  {NoStop}%
\bibitem [{\citenamefont {Esposito}\ \emph {et~al.}(2009)\citenamefont
  {Esposito}, \citenamefont {Harbola},\ and\ \citenamefont
  {Mukamel}}]{Esposito2009}%
  \BibitemOpen
  \bibfield  {author} {\bibinfo {author} {\bibfnamefont {M.}~\bibnamefont
  {Esposito}}, \bibinfo {author} {\bibfnamefont {U.}~\bibnamefont {Harbola}}, \
  and\ \bibinfo {author} {\bibfnamefont {S.}~\bibnamefont {Mukamel}},\ }\href
  {\doibase 10.1103/RevModPhys.81.1665} {\bibfield  {journal} {\bibinfo
  {journal} {Reviews of Modern Physics}\ }\textbf {\bibinfo {volume} {81}},\
  \bibinfo {pages} {1665} (\bibinfo {year} {2009})}\BibitemShut {NoStop}%
\bibitem [{\citenamefont {Brandes}(2008)}]{Brandes2008}%
  \BibitemOpen
  \bibfield  {author} {\bibinfo {author} {\bibfnamefont {T.}~\bibnamefont
  {Brandes}},\ }\href {\doibase 10.1002/andp.200810306} {\bibfield  {journal}
  {\bibinfo  {journal} {Annalen der Physik (Leipzig)}\ }\textbf {\bibinfo
  {volume} {17}},\ \bibinfo {pages} {477} (\bibinfo {year} {2008})}\BibitemShut
  {NoStop}%
\bibitem [{\citenamefont {Touchette}(2009)}]{Touchette2009}%
  \BibitemOpen
  \bibfield  {author} {\bibinfo {author} {\bibfnamefont {H.}~\bibnamefont
  {Touchette}},\ }\href {\doibase 10.1016/j.physrep.2009.05.002} {\bibfield
  {journal} {\bibinfo  {journal} {Physics Reports}\ }\textbf {\bibinfo {volume}
  {478}},\ \bibinfo {pages} {1} (\bibinfo {year} {2009})}\BibitemShut {NoStop}%
\bibitem [{\citenamefont {Touchette}(2012)}]{Touchette2012}%
  \BibitemOpen
  \bibfield  {author} {\bibinfo {author} {\bibfnamefont {H.}~\bibnamefont
  {Touchette}},\ }\href@noop {} {\ ,\ \bibinfo {pages} {1} (\bibinfo {year}
  {2012})}\BibitemShut {NoStop}%
\bibitem [{\citenamefont {Cohen-Tannoudji}\ and\ \citenamefont
  {Dalibard}(1986)}]{Cohen-Tannoudji1986}%
  \BibitemOpen
  \bibfield  {author} {\bibinfo {author} {\bibfnamefont {C.}~\bibnamefont
  {Cohen-Tannoudji}}\ and\ \bibinfo {author} {\bibfnamefont {J.}~\bibnamefont
  {Dalibard}},\ }\href {\doibase 10.1209/0295-5075/1/9/004} {\bibfield
  {journal} {\bibinfo  {journal} {Europhysics Letters (EPL)}\ }\textbf
  {\bibinfo {volume} {1}},\ \bibinfo {pages} {441} (\bibinfo {year}
  {1986})}\BibitemShut {NoStop}%
\bibitem [{\citenamefont {Plenio}\ and\ \citenamefont
  {Knight}(1998)}]{Plenio1998a}%
  \BibitemOpen
  \bibfield  {author} {\bibinfo {author} {\bibfnamefont {M.~B.}\ \bibnamefont
  {Plenio}}\ and\ \bibinfo {author} {\bibfnamefont {P.~L.}\ \bibnamefont
  {Knight}},\ }\href {\doibase 10.1103/RevModPhys.70.101} {\bibfield  {journal}
  {\bibinfo  {journal} {Reviews of Modern Physics}\ }\textbf {\bibinfo {volume}
  {70}},\ \bibinfo {pages} {101} (\bibinfo {year} {1998})}\BibitemShut
  {NoStop}%
\bibitem [{\citenamefont {Schaller}\ \emph {et~al.}(2009)\citenamefont
  {Schaller}, \citenamefont {Kie{\ss}lich},\ and\ \citenamefont
  {Brandes}}]{Schaller2009}%
  \BibitemOpen
  \bibfield  {author} {\bibinfo {author} {\bibfnamefont {G.}~\bibnamefont
  {Schaller}}, \bibinfo {author} {\bibfnamefont {G.}~\bibnamefont
  {Kie{\ss}lich}}, \ and\ \bibinfo {author} {\bibfnamefont {T.}~\bibnamefont
  {Brandes}},\ }\href {\doibase 10.1103/PhysRevB.80.245107} {\bibfield
  {journal} {\bibinfo  {journal} {Physical Review B}\ }\textbf {\bibinfo
  {volume} {80}},\ \bibinfo {pages} {245107} (\bibinfo {year}
  {2009})}\BibitemShut {NoStop}%
\bibitem [{\citenamefont {Albert}\ \emph {et~al.}(2011)\citenamefont {Albert},
  \citenamefont {Flindt},\ and\ \citenamefont {B{\"{u}}ttiker}}]{Albert2011}%
  \BibitemOpen
  \bibfield  {author} {\bibinfo {author} {\bibfnamefont {M.}~\bibnamefont
  {Albert}}, \bibinfo {author} {\bibfnamefont {C.}~\bibnamefont {Flindt}}, \
  and\ \bibinfo {author} {\bibfnamefont {M.}~\bibnamefont {B{\"{u}}ttiker}},\
  }\href {\doibase 10.1103/PhysRevLett.107.086805} {\bibfield  {journal}
  {\bibinfo  {journal} {Physical Review Letters}\ }\textbf {\bibinfo {volume}
  {107}},\ \bibinfo {pages} {086805} (\bibinfo {year} {2011})}\BibitemShut
  {NoStop}%
\bibitem [{\citenamefont {Albert}\ \emph {et~al.}(2012)\citenamefont {Albert},
  \citenamefont {Haack}, \citenamefont {Flindt},\ and\ \citenamefont
  {B{\"{u}}ttiker}}]{Albert2012}%
  \BibitemOpen
  \bibfield  {author} {\bibinfo {author} {\bibfnamefont {M.}~\bibnamefont
  {Albert}}, \bibinfo {author} {\bibfnamefont {G.}~\bibnamefont {Haack}},
  \bibinfo {author} {\bibfnamefont {C.}~\bibnamefont {Flindt}}, \ and\ \bibinfo
  {author} {\bibfnamefont {M.}~\bibnamefont {B{\"{u}}ttiker}},\ }\href
  {\doibase 10.1103/PhysRevLett.108.186806} {\bibfield  {journal} {\bibinfo
  {journal} {Physical Review Letters}\ }\textbf {\bibinfo {volume} {108}},\
  \bibinfo {pages} {186806} (\bibinfo {year} {2012})}\BibitemShut {NoStop}%
\bibitem [{\citenamefont {Rajabi}\ \emph {et~al.}(2013)\citenamefont {Rajabi},
  \citenamefont {P{\"{o}}ltl},\ and\ \citenamefont {Governale}}]{Rajabi2013}%
  \BibitemOpen
  \bibfield  {author} {\bibinfo {author} {\bibfnamefont {L.}~\bibnamefont
  {Rajabi}}, \bibinfo {author} {\bibfnamefont {C.}~\bibnamefont {P{\"{o}}ltl}},
  \ and\ \bibinfo {author} {\bibfnamefont {M.}~\bibnamefont {Governale}},\
  }\href {\doibase 10.1103/PhysRevLett.111.067002} {\bibfield  {journal}
  {\bibinfo  {journal} {Physical Review Letters}\ }\textbf {\bibinfo {volume}
  {111}},\ \bibinfo {pages} {067002} (\bibinfo {year} {2013})}\BibitemShut
  {NoStop}%
\bibitem [{\citenamefont {Thomas}\ and\ \citenamefont
  {Flindt}(2013)}]{Thomas2013}%
  \BibitemOpen
  \bibfield  {author} {\bibinfo {author} {\bibfnamefont {K.~H.}\ \bibnamefont
  {Thomas}}\ and\ \bibinfo {author} {\bibfnamefont {C.}~\bibnamefont
  {Flindt}},\ }\href {\doibase 10.1103/PhysRevB.87.121405} {\bibfield
  {journal} {\bibinfo  {journal} {Physical Review B}\ }\textbf {\bibinfo
  {volume} {87}},\ \bibinfo {pages} {121405} (\bibinfo {year}
  {2013})}\BibitemShut {NoStop}%
\bibitem [{\citenamefont {Thomas}\ and\ \citenamefont
  {Flindt}(2014)}]{Thomas2014}%
  \BibitemOpen
  \bibfield  {author} {\bibinfo {author} {\bibfnamefont {K.~H.}\ \bibnamefont
  {Thomas}}\ and\ \bibinfo {author} {\bibfnamefont {C.}~\bibnamefont
  {Flindt}},\ }\href {\doibase 10.1103/PhysRevB.89.245420} {\bibfield
  {journal} {\bibinfo  {journal} {Physical Review B}\ }\textbf {\bibinfo
  {volume} {89}},\ \bibinfo {pages} {245420} (\bibinfo {year}
  {2014})}\BibitemShut {NoStop}%
\bibitem [{\citenamefont {Haack}\ \emph {et~al.}(2014)\citenamefont {Haack},
  \citenamefont {Albert},\ and\ \citenamefont {Flindt}}]{Haack2014}%
  \BibitemOpen
  \bibfield  {author} {\bibinfo {author} {\bibfnamefont {G.}~\bibnamefont
  {Haack}}, \bibinfo {author} {\bibfnamefont {M.}~\bibnamefont {Albert}}, \
  and\ \bibinfo {author} {\bibfnamefont {C.}~\bibnamefont {Flindt}},\ }\href
  {\doibase 10.1103/PhysRevB.90.205429} {\bibfield  {journal} {\bibinfo
  {journal} {Physical Review B}\ }\textbf {\bibinfo {volume} {90}},\ \bibinfo
  {pages} {205429} (\bibinfo {year} {2014})}\BibitemShut {NoStop}%
\bibitem [{\citenamefont {Dasenbrook}\ \emph {et~al.}(2015)\citenamefont
  {Dasenbrook}, \citenamefont {Hofer},\ and\ \citenamefont
  {Flindt}}]{Dasenbrook2015}%
  \BibitemOpen
  \bibfield  {author} {\bibinfo {author} {\bibfnamefont {D.}~\bibnamefont
  {Dasenbrook}}, \bibinfo {author} {\bibfnamefont {P.~P.}\ \bibnamefont
  {Hofer}}, \ and\ \bibinfo {author} {\bibfnamefont {C.}~\bibnamefont
  {Flindt}},\ }\href {\doibase 10.1103/PhysRevB.91.195420} {\bibfield
  {journal} {\bibinfo  {journal} {Physical Review B}\ }\textbf {\bibinfo
  {volume} {91}},\ \bibinfo {pages} {195420} (\bibinfo {year}
  {2015})}\BibitemShut {NoStop}%
\bibitem [{\citenamefont
  {Ptaszy{\'{n}}ski}(2017{\natexlab{a}})}]{Ptaszynski2017a}%
  \BibitemOpen
  \bibfield  {author} {\bibinfo {author} {\bibfnamefont {K.}~\bibnamefont
  {Ptaszy{\'{n}}ski}},\ }\href {\doibase 10.1103/PhysRevB.96.035409} {\bibfield
   {journal} {\bibinfo  {journal} {Physical Review B}\ }\textbf {\bibinfo
  {volume} {96}},\ \bibinfo {pages} {035409} (\bibinfo {year}
  {2017}{\natexlab{a}})}\BibitemShut {NoStop}%
\bibitem [{\citenamefont
  {Ptaszy{\'{n}}ski}(2017{\natexlab{b}})}]{Ptaszynski2017}%
  \BibitemOpen
  \bibfield  {author} {\bibinfo {author} {\bibfnamefont {K.}~\bibnamefont
  {Ptaszy{\'{n}}ski}},\ }\href {\doibase 10.1103/PhysRevB.95.045306} {\bibfield
   {journal} {\bibinfo  {journal} {Physical Review B}\ }\textbf {\bibinfo
  {volume} {95}},\ \bibinfo {pages} {045306} (\bibinfo {year}
  {2017}{\natexlab{b}})}\BibitemShut {NoStop}%
\bibitem [{\citenamefont {Stegmann}\ \emph {et~al.}(2021)\citenamefont
  {Stegmann}, \citenamefont {Sothmann}, \citenamefont {K{\"{o}}nig},\ and\
  \citenamefont {Flindt}}]{Stegmann2021}%
  \BibitemOpen
  \bibfield  {author} {\bibinfo {author} {\bibfnamefont {P.}~\bibnamefont
  {Stegmann}}, \bibinfo {author} {\bibfnamefont {B.}~\bibnamefont {Sothmann}},
  \bibinfo {author} {\bibfnamefont {J.}~\bibnamefont {K{\"{o}}nig}}, \ and\
  \bibinfo {author} {\bibfnamefont {C.}~\bibnamefont {Flindt}},\ }\href
  {\doibase 10.1103/PhysRevLett.127.096803} {\bibfield  {journal} {\bibinfo
  {journal} {Physical Review Letters}\ }\textbf {\bibinfo {volume} {127}},\
  \bibinfo {pages} {096803} (\bibinfo {year} {2021})}\BibitemShut {NoStop}%
\bibitem [{\citenamefont {Stegmann}\ \emph {et~al.}(2018)\citenamefont
  {Stegmann}, \citenamefont {K{\"{o}}nig},\ and\ \citenamefont
  {Weiss}}]{Stegmann2018}%
  \BibitemOpen
  \bibfield  {author} {\bibinfo {author} {\bibfnamefont {P.}~\bibnamefont
  {Stegmann}}, \bibinfo {author} {\bibfnamefont {J.}~\bibnamefont
  {K{\"{o}}nig}}, \ and\ \bibinfo {author} {\bibfnamefont {S.}~\bibnamefont
  {Weiss}},\ }\href {\doibase 10.1103/PhysRevB.98.035409} {\bibfield  {journal}
  {\bibinfo  {journal} {Physical Review B}\ }\textbf {\bibinfo {volume} {98}},\
  \bibinfo {pages} {035409} (\bibinfo {year} {2018})}\BibitemShut {NoStop}%
\bibitem [{\citenamefont {Walldorf}\ \emph {et~al.}(2018)\citenamefont
  {Walldorf}, \citenamefont {Padurariu}, \citenamefont {Jauho},\ and\
  \citenamefont {Flindt}}]{Walldorf2018}%
  \BibitemOpen
  \bibfield  {author} {\bibinfo {author} {\bibfnamefont {N.}~\bibnamefont
  {Walldorf}}, \bibinfo {author} {\bibfnamefont {C.}~\bibnamefont {Padurariu}},
  \bibinfo {author} {\bibfnamefont {A.-P.}\ \bibnamefont {Jauho}}, \ and\
  \bibinfo {author} {\bibfnamefont {C.}~\bibnamefont {Flindt}},\ }\href
  {\doibase 10.1103/PhysRevLett.120.087701} {\bibfield  {journal} {\bibinfo
  {journal} {Physical Review Letters}\ }\textbf {\bibinfo {volume} {120}},\
  \bibinfo {pages} {087701} (\bibinfo {year} {2018})}\BibitemShut {NoStop}%
\bibitem [{\citenamefont {Kleinherbers}\ \emph {et~al.}(2021)\citenamefont
  {Kleinherbers}, \citenamefont {Stegmann},\ and\ \citenamefont
  {K{\"{o}}nig}}]{Kleinherbers2021}%
  \BibitemOpen
  \bibfield  {author} {\bibinfo {author} {\bibfnamefont {E.}~\bibnamefont
  {Kleinherbers}}, \bibinfo {author} {\bibfnamefont {P.}~\bibnamefont
  {Stegmann}}, \ and\ \bibinfo {author} {\bibfnamefont {J.}~\bibnamefont
  {K{\"{o}}nig}},\ }\href {\doibase 10.1103/PhysRevB.104.165304} {\bibfield
  {journal} {\bibinfo  {journal} {Physical Review B}\ }\textbf {\bibinfo
  {volume} {104}},\ \bibinfo {pages} {165304} (\bibinfo {year}
  {2021})}\BibitemShut {NoStop}%
\bibitem [{\citenamefont {Wiseman}\ and\ \citenamefont
  {Milburn}(2009)}]{Wiseman2009}%
  \BibitemOpen
  \bibfield  {author} {\bibinfo {author} {\bibfnamefont {H.~M.}\ \bibnamefont
  {Wiseman}}\ and\ \bibinfo {author} {\bibfnamefont {G.~J.}\ \bibnamefont
  {Milburn}},\ }\href@noop {} {\emph {\bibinfo {title} {{Quantum measurement
  and control}}}}\ (\bibinfo  {publisher} {Cambridge University Press},\
  \bibinfo {address} {New York},\ \bibinfo {year} {2009})\BibitemShut {NoStop}%
\bibitem [{\citenamefont {Bruus}\ and\ \citenamefont
  {Flensberg}(2004)}]{Logan2005}%
  \BibitemOpen
  \bibfield  {author} {\bibinfo {author} {\bibfnamefont {H.}~\bibnamefont
  {Bruus}}\ and\ \bibinfo {author} {\bibfnamefont {K.}~\bibnamefont
  {Flensberg}},\ }\href {\doibase 10.1088/0305-4470/38/8/B01} {\emph {\bibinfo
  {title} {{Many-Body Quantum Theory in Condensed Matter Physics}}}}\ (\bibinfo
   {publisher} {Oxford University Press},\ \bibinfo {year} {2004})\ p.\
  \bibinfo {pages} {466}\BibitemShut {NoStop}%
\bibitem [{\citenamefont {Blankenbecler}\ \emph {et~al.}(1981)\citenamefont
  {Blankenbecler}, \citenamefont {Scalapino},\ and\ \citenamefont
  {Sugar}}]{Blankenbecler1981}%
  \BibitemOpen
  \bibfield  {author} {\bibinfo {author} {\bibfnamefont {R.}~\bibnamefont
  {Blankenbecler}}, \bibinfo {author} {\bibfnamefont {D.~J.}\ \bibnamefont
  {Scalapino}}, \ and\ \bibinfo {author} {\bibfnamefont {R.~L.}\ \bibnamefont
  {Sugar}},\ }\href {\doibase 10.1103/PhysRevD.24.2278} {\bibfield  {journal}
  {\bibinfo  {journal} {Physical Review D}\ }\textbf {\bibinfo {volume} {24}},\
  \bibinfo {pages} {2278} (\bibinfo {year} {1981})}\BibitemShut {NoStop}%
\bibitem [{\citenamefont {Klich}(2014)}]{Klich2014}%
  \BibitemOpen
  \bibfield  {author} {\bibinfo {author} {\bibfnamefont {I.}~\bibnamefont
  {Klich}},\ }\href {\doibase 10.1088/1742-5468/2014/11/P11006} {\bibfield
  {journal} {\bibinfo  {journal} {Journal of Statistical Mechanics: Theory and
  Experiment}\ }\textbf {\bibinfo {volume} {2014}},\ \bibinfo {pages} {1}
  (\bibinfo {year} {2014})}\BibitemShut {NoStop}%
\bibitem [{\citenamefont {Karevski}\ and\ \citenamefont
  {Platini}(2009)}]{Karevski2009}%
  \BibitemOpen
  \bibfield  {author} {\bibinfo {author} {\bibfnamefont {D.}~\bibnamefont
  {Karevski}}\ and\ \bibinfo {author} {\bibfnamefont {T.}~\bibnamefont
  {Platini}},\ }\href {\doibase 10.1103/PhysRevLett.102.207207} {\bibfield
  {journal} {\bibinfo  {journal} {Physical Review Letters}\ }\textbf {\bibinfo
  {volume} {102}},\ \bibinfo {pages} {207207} (\bibinfo {year}
  {2009})}\BibitemShut {NoStop}%
\bibitem [{\citenamefont {{\v{Z}}nidari{\v{c}}}(2010)}]{Znidaric2010a}%
  \BibitemOpen
  \bibfield  {author} {\bibinfo {author} {\bibfnamefont {M.}~\bibnamefont
  {{\v{Z}}nidari{\v{c}}}},\ }\href {\doibase 10.1088/1742-5468/2010/05/L05002}
  {\bibfield  {journal} {\bibinfo  {journal} {Journal of Statistical Mechanics:
  Theory and Experiment}\ }\textbf {\bibinfo {volume} {L05002}},\ \bibinfo
  {pages} {1742} (\bibinfo {year} {2010})}\BibitemShut {NoStop}%
\bibitem [{\citenamefont {Asadian}\ \emph {et~al.}(2013)\citenamefont
  {Asadian}, \citenamefont {Manzano}, \citenamefont {Tiersch},\ and\
  \citenamefont {Briegel}}]{Asadian2013}%
  \BibitemOpen
  \bibfield  {author} {\bibinfo {author} {\bibfnamefont {A.}~\bibnamefont
  {Asadian}}, \bibinfo {author} {\bibfnamefont {D.}~\bibnamefont {Manzano}},
  \bibinfo {author} {\bibfnamefont {M.}~\bibnamefont {Tiersch}}, \ and\
  \bibinfo {author} {\bibfnamefont {H.~J.}\ \bibnamefont {Briegel}},\ }\href
  {\doibase 10.1103/PhysRevE.87.012109} {\bibfield  {journal} {\bibinfo
  {journal} {Physical Review E}\ }\textbf {\bibinfo {volume} {87}},\ \bibinfo
  {pages} {012109} (\bibinfo {year} {2013})}\BibitemShut {NoStop}%
\bibitem [{Note1()}]{Note1}%
  \BibitemOpen
  \bibinfo {note} {{\protect \color {black}This analogy is limited by the fact
  that $P(t,1_+|1_+) = 0$ for $t = 0$.}}\BibitemShut {Stop}%
\bibitem [{\citenamefont {Mascarenhas}\ \emph {et~al.}(2015)\citenamefont
  {Mascarenhas}, \citenamefont {Flayac},\ and\ \citenamefont
  {Savona}}]{Mascarenhas2015}%
  \BibitemOpen
  \bibfield  {author} {\bibinfo {author} {\bibfnamefont {E.}~\bibnamefont
  {Mascarenhas}}, \bibinfo {author} {\bibfnamefont {H.}~\bibnamefont {Flayac}},
  \ and\ \bibinfo {author} {\bibfnamefont {V.}~\bibnamefont {Savona}},\ }\href
  {\doibase 10.1103/PhysRevA.92.022116} {\bibfield  {journal} {\bibinfo
  {journal} {Physical Review A - Atomic, Molecular, and Optical Physics}\
  }\textbf {\bibinfo {volume} {92}},\ \bibinfo {pages} {022116} (\bibinfo
  {year} {2015})}\BibitemShut {NoStop}%
\bibitem [{\citenamefont {Cui}\ \emph {et~al.}(2015)\citenamefont {Cui},
  \citenamefont {Cirac},\ and\ \citenamefont {Ba{\~{n}}uls}}]{Cui2015}%
  \BibitemOpen
  \bibfield  {author} {\bibinfo {author} {\bibfnamefont {J.}~\bibnamefont
  {Cui}}, \bibinfo {author} {\bibfnamefont {J.~I.}\ \bibnamefont {Cirac}}, \
  and\ \bibinfo {author} {\bibfnamefont {M.~C.}\ \bibnamefont {Ba{\~{n}}uls}},\
  }\href {\doibase 10.1103/PhysRevLett.114.220601} {\bibfield  {journal}
  {\bibinfo  {journal} {Physical Review Letters}\ }\textbf {\bibinfo {volume}
  {114}},\ \bibinfo {pages} {220601} (\bibinfo {year} {2015})}\BibitemShut
  {NoStop}%
\bibitem [{\citenamefont {Werner}\ \emph {et~al.}(2016)\citenamefont {Werner},
  \citenamefont {Jaschke}, \citenamefont {Silvi}, \citenamefont {Calarco},
  \citenamefont {Eisert},\ and\ \citenamefont {Montangero}}]{Werner2014}%
  \BibitemOpen
  \bibfield  {author} {\bibinfo {author} {\bibfnamefont {A.~H.}\ \bibnamefont
  {Werner}}, \bibinfo {author} {\bibfnamefont {D.}~\bibnamefont {Jaschke}},
  \bibinfo {author} {\bibfnamefont {P.}~\bibnamefont {Silvi}}, \bibinfo
  {author} {\bibfnamefont {T.}~\bibnamefont {Calarco}}, \bibinfo {author}
  {\bibfnamefont {J.}~\bibnamefont {Eisert}}, \ and\ \bibinfo {author}
  {\bibfnamefont {S.}~\bibnamefont {Montangero}},\ }\href {\doibase
  10.1103/PhysRevLett.116.237201} {\bibfield  {journal} {\bibinfo  {journal}
  {Physical Review Letters}\ }\textbf {\bibinfo {volume} {116}},\ \bibinfo
  {pages} {237201} (\bibinfo {year} {2016})}\BibitemShut {NoStop}%
\bibitem [{\citenamefont {Jaschke}\ \emph {et~al.}(2018)\citenamefont
  {Jaschke}, \citenamefont {Montangero},\ and\ \citenamefont
  {Carr}}]{Jaschke2018}%
  \BibitemOpen
  \bibfield  {author} {\bibinfo {author} {\bibfnamefont {D.}~\bibnamefont
  {Jaschke}}, \bibinfo {author} {\bibfnamefont {S.}~\bibnamefont {Montangero}},
  \ and\ \bibinfo {author} {\bibfnamefont {L.~D.}\ \bibnamefont {Carr}},\
  }\href {http://arxiv.org/abs/1804.09796} {\ ,\ \bibinfo {pages} {1} (\bibinfo
  {year} {2018})}\BibitemShut {NoStop}%
\bibitem [{\citenamefont {Brenes}\ \emph {et~al.}(2020)\citenamefont {Brenes},
  \citenamefont {Mendoza-Arenas}, \citenamefont {Purkayastha}, \citenamefont
  {Mitchison}, \citenamefont {Clark},\ and\ \citenamefont
  {Goold}}]{Brenes2020}%
  \BibitemOpen
  \bibfield  {author} {\bibinfo {author} {\bibfnamefont {M.}~\bibnamefont
  {Brenes}}, \bibinfo {author} {\bibfnamefont {J.~J.}\ \bibnamefont
  {Mendoza-Arenas}}, \bibinfo {author} {\bibfnamefont {A.}~\bibnamefont
  {Purkayastha}}, \bibinfo {author} {\bibfnamefont {M.~T.}\ \bibnamefont
  {Mitchison}}, \bibinfo {author} {\bibfnamefont {S.~R.}\ \bibnamefont
  {Clark}}, \ and\ \bibinfo {author} {\bibfnamefont {J.}~\bibnamefont
  {Goold}},\ }\href {\doibase 10.1103/PhysRevX.10.031040} {\bibfield  {journal}
  {\bibinfo  {journal} {Physical Review X}\ }\textbf {\bibinfo {volume} {10}},\
  \bibinfo {pages} {031040} (\bibinfo {year} {2020})}\BibitemShut {NoStop}%
\bibitem [{\citenamefont {Casagrande}\ \emph {et~al.}(2021)\citenamefont
  {Casagrande}, \citenamefont {Poletti},\ and\ \citenamefont
  {Landi}}]{Casagrande2020}%
  \BibitemOpen
  \bibfield  {author} {\bibinfo {author} {\bibfnamefont {H.~P.}\ \bibnamefont
  {Casagrande}}, \bibinfo {author} {\bibfnamefont {D.}~\bibnamefont {Poletti}},
  \ and\ \bibinfo {author} {\bibfnamefont {G.~T.}\ \bibnamefont {Landi}},\
  }\href {\doibase 10.1016/j.cpc.2021.108060} {\bibfield  {journal} {\bibinfo
  {journal} {Computer Physics Communications}\ }\textbf {\bibinfo {volume}
  {267}},\ \bibinfo {pages} {108060} (\bibinfo {year} {2021})}\BibitemShut
  {NoStop}%
\bibitem [{\citenamefont {Varma}\ \emph {et~al.}(2017)\citenamefont {Varma},
  \citenamefont {de~Mulatier},\ and\ \citenamefont
  {{\v{Z}}nidari{\v{c}}}}]{Varma2017}%
  \BibitemOpen
  \bibfield  {author} {\bibinfo {author} {\bibfnamefont {V.~K.}\ \bibnamefont
  {Varma}}, \bibinfo {author} {\bibfnamefont {C.}~\bibnamefont {de~Mulatier}},
  \ and\ \bibinfo {author} {\bibfnamefont {M.}~\bibnamefont
  {{\v{Z}}nidari{\v{c}}}},\ }\href {\doibase 10.1103/PhysRevE.96.032130}
  {\bibfield  {journal} {\bibinfo  {journal} {Physical Review E}\ }\textbf
  {\bibinfo {volume} {96}},\ \bibinfo {pages} {032130} (\bibinfo {year}
  {2017})}\BibitemShut {NoStop}%
\bibitem [{\citenamefont {{\v{Z}}nidari{\v{c}}}\ \emph
  {et~al.}(2017)\citenamefont {{\v{Z}}nidari{\v{c}}}, \citenamefont
  {Mendoza‐Arenas}, \citenamefont {Clark},\ and\ \citenamefont
  {Goold}}]{Znidaric2017}%
  \BibitemOpen
  \bibfield  {author} {\bibinfo {author} {\bibfnamefont {M.}~\bibnamefont
  {{\v{Z}}nidari{\v{c}}}}, \bibinfo {author} {\bibfnamefont {J.~J.}\
  \bibnamefont {Mendoza‐Arenas}}, \bibinfo {author} {\bibfnamefont {S.~R.}\
  \bibnamefont {Clark}}, \ and\ \bibinfo {author} {\bibfnamefont
  {J.}~\bibnamefont {Goold}},\ }\href {\doibase 10.1002/andp.201600298}
  {\bibfield  {journal} {\bibinfo  {journal} {Annalen der Physik}\ }\textbf
  {\bibinfo {volume} {529}},\ \bibinfo {pages} {1600298} (\bibinfo {year}
  {2017})}\BibitemShut {NoStop}%
\bibitem [{\citenamefont {Ganeshan}\ \emph {et~al.}(2015)\citenamefont
  {Ganeshan}, \citenamefont {Pixley},\ and\ \citenamefont
  {Das~Sarma}}]{Ganeshan2015}%
  \BibitemOpen
  \bibfield  {author} {\bibinfo {author} {\bibfnamefont {S.}~\bibnamefont
  {Ganeshan}}, \bibinfo {author} {\bibfnamefont {J.~H.}\ \bibnamefont
  {Pixley}}, \ and\ \bibinfo {author} {\bibfnamefont {S.}~\bibnamefont
  {Das~Sarma}},\ }\href {\doibase 10.1103/PhysRevLett.114.146601} {\bibfield
  {journal} {\bibinfo  {journal} {Physical Review Letters}\ }\textbf {\bibinfo
  {volume} {114}},\ \bibinfo {pages} {146601} (\bibinfo {year}
  {2015})}\BibitemShut {NoStop}%
\bibitem [{\citenamefont {{\v{Z}}nidari{\v{c}}}\ and\ \citenamefont
  {Horvat}(2013)}]{Znidaric2013a}%
  \BibitemOpen
  \bibfield  {author} {\bibinfo {author} {\bibfnamefont {M.}~\bibnamefont
  {{\v{Z}}nidari{\v{c}}}}\ and\ \bibinfo {author} {\bibfnamefont
  {M.}~\bibnamefont {Horvat}},\ }\href {\doibase 10.1140/epjb/e2012-30730-9}
  {\bibfield  {journal} {\bibinfo  {journal} {The European Physical Journal B}\
  }\textbf {\bibinfo {volume} {86}},\ \bibinfo {pages} {67} (\bibinfo {year}
  {2013})}\BibitemShut {NoStop}%
\bibitem [{\citenamefont {Hiramoto}\ and\ \citenamefont
  {Kohmoto}(1992)}]{Hiramoto1992}%
  \BibitemOpen
  \bibfield  {author} {\bibinfo {author} {\bibfnamefont {H.}~\bibnamefont
  {Hiramoto}}\ and\ \bibinfo {author} {\bibfnamefont {M.}~\bibnamefont
  {Kohmoto}},\ }\href@noop {} {\bibfield  {journal} {\bibinfo  {journal}
  {International Journal of Modern Physics B}\ }\textbf {\bibinfo {volume}
  {6}},\ \bibinfo {pages} {281} (\bibinfo {year} {1992})}\BibitemShut {NoStop}%
\bibitem [{\citenamefont {Purkayastha}\ \emph {et~al.}(2017)\citenamefont
  {Purkayastha}, \citenamefont {Dhar},\ and\ \citenamefont
  {Kulkarni}}]{Purkayastha2017a}%
  \BibitemOpen
  \bibfield  {author} {\bibinfo {author} {\bibfnamefont {A.}~\bibnamefont
  {Purkayastha}}, \bibinfo {author} {\bibfnamefont {A.}~\bibnamefont {Dhar}}, \
  and\ \bibinfo {author} {\bibfnamefont {M.}~\bibnamefont {Kulkarni}},\ }\href
  {\doibase 10.1103/PhysRevB.96.180204} {\bibfield  {journal} {\bibinfo
  {journal} {Physical Review B}\ }\textbf {\bibinfo {volume} {96}},\ \bibinfo
  {pages} {180204} (\bibinfo {year} {2017})}\BibitemShut {NoStop}%
\bibitem [{\citenamefont {Chiaracane}\ \emph {et~al.}(2019)\citenamefont
  {Chiaracane}, \citenamefont {Mitchison}, \citenamefont {Purkayastha},
  \citenamefont {Haack},\ and\ \citenamefont {Goold}}]{Chiaracane2019}%
  \BibitemOpen
  \bibfield  {author} {\bibinfo {author} {\bibfnamefont {C.}~\bibnamefont
  {Chiaracane}}, \bibinfo {author} {\bibfnamefont {M.~T.}\ \bibnamefont
  {Mitchison}}, \bibinfo {author} {\bibfnamefont {A.}~\bibnamefont
  {Purkayastha}}, \bibinfo {author} {\bibfnamefont {G.}~\bibnamefont {Haack}},
  \ and\ \bibinfo {author} {\bibfnamefont {J.}~\bibnamefont {Goold}},\ }\href
  {http://arxiv.org/abs/1908.05139} {\ ,\ \bibinfo {pages} {1} (\bibinfo {year}
  {2019})}\BibitemShut {NoStop}%
\bibitem [{\citenamefont {Varma}\ and\ \citenamefont
  {{\v{Z}}nidari{\v{c}}}(2019)}]{Varma2019}%
  \BibitemOpen
  \bibfield  {author} {\bibinfo {author} {\bibfnamefont {V.~K.}\ \bibnamefont
  {Varma}}\ and\ \bibinfo {author} {\bibfnamefont {M.}~\bibnamefont
  {{\v{Z}}nidari{\v{c}}}},\ }\href {http://arxiv.org/abs/1905.03128} {\ \textbf
  {\bibinfo {volume} {2}},\ \bibinfo {pages} {1} (\bibinfo {year}
  {2019})}\BibitemShut {NoStop}%
\bibitem [{\citenamefont {Lacerda}\ \emph {et~al.}(2021)\citenamefont
  {Lacerda}, \citenamefont {Goold},\ and\ \citenamefont {Landi}}]{Lacerda2021}%
  \BibitemOpen
  \bibfield  {author} {\bibinfo {author} {\bibfnamefont {A.~M.}\ \bibnamefont
  {Lacerda}}, \bibinfo {author} {\bibfnamefont {J.}~\bibnamefont {Goold}}, \
  and\ \bibinfo {author} {\bibfnamefont {G.~T.}\ \bibnamefont {Landi}},\ }\href
  {http://arxiv.org/abs/2106.11406} {\ ,\ \bibinfo {pages} {1} (\bibinfo {year}
  {2021})}\BibitemShut {NoStop}%
\bibitem [{\citenamefont {Aspelmeyer}\ \emph {et~al.}(2014)\citenamefont
  {Aspelmeyer}, \citenamefont {Kippenberg},\ and\ \citenamefont
  {Marquardt}}]{Aspelmeyer2014}%
  \BibitemOpen
  \bibfield  {author} {\bibinfo {author} {\bibfnamefont {M.}~\bibnamefont
  {Aspelmeyer}}, \bibinfo {author} {\bibfnamefont {T.~J.}\ \bibnamefont
  {Kippenberg}}, \ and\ \bibinfo {author} {\bibfnamefont {F.}~\bibnamefont
  {Marquardt}},\ }\href {\doibase 10.1103/RevModPhys.86.1391} {\bibfield
  {journal} {\bibinfo  {journal} {Reviews of Modern Physics}\ }\textbf
  {\bibinfo {volume} {86}},\ \bibinfo {pages} {1391} (\bibinfo {year}
  {2014})}\BibitemShut {NoStop}%
\bibitem [{\citenamefont {Bowen}\ and\ \citenamefont
  {Milburn}(2016)}]{Bowen2016}%
  \BibitemOpen
  \bibfield  {author} {\bibinfo {author} {\bibfnamefont {P.~W.}\ \bibnamefont
  {Bowen}}\ and\ \bibinfo {author} {\bibfnamefont {G.~J.}\ \bibnamefont
  {Milburn}},\ }\href@noop {} {\emph {\bibinfo {title} {{Quantum
  Optomechanics}}}}\ (\bibinfo  {publisher} {CRC Press},\ \bibinfo {year}
  {2016})\BibitemShut {NoStop}%
\bibitem [{\citenamefont {Prosen}(2008)}]{Prosen2008}%
  \BibitemOpen
  \bibfield  {author} {\bibinfo {author} {\bibfnamefont {T.}~\bibnamefont
  {Prosen}},\ }\href {\doibase 10.1088/1367-2630/10/4/043026} {\bibfield
  {journal} {\bibinfo  {journal} {New Journal of Physics}\ }\textbf {\bibinfo
  {volume} {10}},\ \bibinfo {pages} {043026} (\bibinfo {year}
  {2008})}\BibitemShut {NoStop}%
\bibitem [{Note2()}]{Note2}%
  \BibitemOpen
  \bibinfo {note} {If one is interested in $i'\neq i$, then the decomposition
  will have the form $e^{\alpha c_i^\dagger c_{i'}} = 1 + \alpha c_i^\dagger
  c_{i'}$.}\BibitemShut {Stop}%
\end{thebibliography}%

\end{document}